\documentclass[lettersize,journal]{IEEEtran}
\usepackage{graphicx}%
\usepackage{multirow}%
\usepackage{mathrsfs}%
\usepackage{xcolor}%
\usepackage{textcomp}%
\usepackage{manyfoot}%
\usepackage{booktabs}%
\usepackage[ruled,lined,linesnumbered]{algorithm2e}
\usepackage{listings}%
\usepackage{pgfplots}
\usepackage{booktabs}
\usepackage{adjustbox}
\usepackage{subfigure} 
\usepackage{float}
\usepackage{booktabs} 
\usepackage{makecell}
\usepackage{caption}
\usepackage{hyperref}
\usepackage{tikz}
\usepackage{colortbl}  
\usepackage{marvosym}
\definecolor{myblue}{RGB}{86,117,145}
\definecolor{mygreen}{RGB}{66,169,138}
\definecolor{myred}{RGB}{143,190,109}
\definecolor{mypurple}{RGB}{249,197,82}
\definecolor{myorange}{RGB}{250,149,29}
\definecolor{mybrown}{RGB}{140, 86, 75}
\usetikzlibrary{quantikz}
\usetikzlibrary{backgrounds}  
\usetikzlibrary{arrows,arrows.meta}
\usepackage[numbers,sort&compress]{natbib}
\pgfmathsetmacro{\transparency}{0.3}
\DeclareCaptionLabelSeparator{period}{. }
\begin{document}

\title{Circuit Partitioning and Transmission Cost Optimization in Distributed Quantum Circuits}

\author{Xinyu~Chen,
        Zilu~Chen,
        Pengcheng~Zhu,
        Xueyun~Cheng,
        and~Zhijin~Guan 
        \thanks{Xinyu Chen, Zilu Chen, Xueyun Cheng and Zhijin Guan, School of Artificial Intelligence and Computer Science, Nantong University, Nantong, 226019, China, E-mail: xinyu\_chen@stmail.ntu.edu.cn, zilu\_chen@stmail.ntu.edu.cn, chen.xy@ntu.edu.cn, guan.zj@ntu.edu.cn.}%
        \thanks{Pengcheng Zhu, College of Information Engineering, Taizhou University, Taizhou, 225300, China, E-mail: zhupcnt@163.com.}%
}

\markboth{}%
{Shell \MakeLowercase{\textit{et al.}}: A Sample Article Using IEEEtran.cls for IEEE Journals}


\maketitle

\begin{abstract}
Given the limitations on the number of qubits in current noisy intermediate-scale quantum (NISQ) devices, the implementation of large-scale quantum algorithms on such devices is challenging, prompting research into distributed quantum computing. This paper focuses on the issue of excessive communication complexity in distributed quantum computing based on the quantum circuit model. To reduce the number of quantum state transmissions, i.e., the transmission cost, in distributed quantum circuits, a circuit partitioning method based on the Quadratic Unconstrained Binary Optimization (QUBO) model is proposed, coupled with the lookahead method for transmission cost optimization. Initially, the problem of distributed quantum circuit partitioning is transformed into a graph minimum cut problem. The QUBO model, which can be accelerated by quantum annealing algorithms, is introduced to minimize the number of quantum gates between quantum processing units (QPUs) and the transmission cost. Subsequently, the dynamic lookahead strategy for the selection of transmission qubits is proposed to optimize the transmission cost in distributed quantum circuits. Finally, through numerical simulations, the impact of different circuit partitioning indicators on the transmission cost is explored, and the proposed method is evaluated on benchmark circuits. Experimental results demonstrate that the proposed circuit partitioning method has a shorter runtime compared with current circuit partitioning methods. Additionally, the transmission cost optimized by the proposed method is significantly lower than that of current transmission cost optimization methods, achieving noticeable improvements across different numbers of partitions.
\end{abstract}


\section{Introduction}
\IEEEPARstart{W}{ith} the release of noisy intermediate-scale quantum (NISQ) chips by IBM and Google \cite{arute2019quantum,jurcevic2021demonstration}, quantum algorithm compilation for NISQ devices has gained notable attention. However, the limited number of qubits available in current NISQ devices cannot meet the requirements of large-scale quantum algorithms. Quantum algorithms that can solve practical problems, due to their high qubit count, cannot be executed on current NISQ devices. To address the execution of quantum algorithms at the scale of tens of thousands or even millions of qubits, distributed quantum computing is considered as the key to this challenge.

Distributed quantum computing aims to decompose a large problem or circuit into multiple parts and distribute them across several quantum processing units (QPUs) for processing. This requires establishing classical or quantum channels between multiple QPUs to ensure quantum information exchange between subcircuits. According to the architecture diagram of distributed quantum computing~\cite{cuomo2020towards}, this architecture can be broadly divided into two layers: hardware and software. The hardware layer consists of network facilities and computing nodes, while the software layer includes distributed quantum algorithms and their compilation. In the software layer, distributed quantum algorithms decompose a specific computational problem into multiple subproblems and design quantum algorithms for these subproblems to achieve distributed quantum computing. Currently implemented distributed quantum algorithms include distributed Grover's algorithm \cite{qiu2024distributed,zhou2023distributed}, distributed Shor's algorithm~\cite{Xiao2023,xiao2023distributed}, etc. Distributed quantum circuit compilation involves partitioning quantum circuits that cannot be executed on a single QPU and mapping the subcircuits to multiple QPUs. Due to its strong generality and good scalability, distributed quantum circuit compilation has attracted significant attention from researchers.

There are multiple technical routes for distributed quantum circuit compilation, such as circuit cutting~\cite{tang2021cutqc,tang2024distributed}, quantum state transmission, and non-local quantum gate operations~\cite{jiang2007distributed, yimsiriwattana2004generalized}. Although circuit cutting-based distributed quantum computing schemes require only classical communication, their sampling complexity grows exponentially and is therefore not considered in this paper. In recent years, quantum communication technologies based on cat entanglement \cite{vlastakis2015characterizing}, quantum teleportation \cite{pirandola2015advances}, and direct quantum state transmission based on quantum interconnects \cite{awschalom2021development} have made significant progress, enabling quantum gate and state transmission between QPUs. However, quantum communication technologies are limited by factors such as qubit stability, qubit transmission loss, and environmental noise. These limitations suggest that excessive quantum communication can detrimentally affect the overall performance of distributed systems. Therefore, reducing the frequency of quantum communications within distributed quantum circuits can help enhance performance indicators such as latency and fidelity.

To reduce the frequency of quantum communication within distributed quantum circuits, Refs. \cite{andres2019automated,daei2020optimized,davarzani2020dynamic,sunkel2023applying,ferrari2021compiler,dadkhah2021new,houshmand2020evolutionary,zomorodi2018optimizing,daei2021improving,ghodsollahee2021connectivity,wu2022autocomm,davarzani2022hierarchical,cheng2023optimization,mao2023qubit,ovide2023mapping,bandic2023mapping} have proposed a range of solutions, focusing on aspects such as circuit partitioning, transmission scheduling, and mapping. In the partitioning of distributed quantum circuits, the references primarily employ methods like hypergraph partitioning \cite{andres2019automated}, graph partitioning based on the Kernighan-Lin (KL) algorithm \cite{daei2020optimized}, bipartite graph partitioning \cite{davarzani2020dynamic}, and qubit partitioning based on genetic algorithms \cite{sunkel2023applying} to reduce the number of remote quantum gates between QPUs, thereby optimizing the frequency of quantum communication. In early works \cite{andres2019automated,ferrari2021compiler,dadkhah2021new}, each remote two-qubit gate between QPUs required independent quantum communication, leading to high communication cost. However, remote two-qubit gates in distributed quantum circuits could be coordinated through one or two communication calls. To this end, Refs. \cite{zomorodi2018optimizing,daei2021improving,ghodsollahee2021connectivity,wu2022autocomm,davarzani2022hierarchical,cheng2023optimization} employ methods such as quantum state and quantum gate transmission scheduling to realize multiple remote two-qubit gates through one or two communication calls, reducing the frequency of quantum communication. In the mapping of distributed quantum circuits, Refs. \cite{mao2023qubit,ovide2023mapping,bandic2023mapping} use parameters such as the number of quantum communications and entangled qubits as optimization indicators to map quantum circuits onto distributed quantum architectures.

This work focuses on direct quantum state transmission as the technology for transferring quantum states in the distributed quantum circuits. It considers the number of quantum state transmissions as the transmission cost and the optimization target. The main contributions of this work are as follows:

\begin{itemize}
\item The distributed quantum circuit partitioning problem can be converted into the minimum cut problem, and the cost indicator for circuit partitioning is optimized. The Quadratic Unconstrained Binary Optimization (QUBO) model that can utilize quantum algorithms for acceleration is proposed for partitioning quantum circuits, aiming to minimize the number of quantum gates that need to be transferred and the transmission cost.
\item The impact of transmission qubits of the quantum gate on the transmission cost is analyzed, defining three types of impact relationships. Based on the impact of different lookahead windows on the transmission cost of distributed circuits, a dynamic lookahead strategy for selecting transmission qubits is proposed to optimize the transmission cost.
\item The transmission cost of distributed quantum circuits is simulated through experimental numerical methods, and the impact of different partitioning indicators on the transmission cost is explored. Comparative experimental results validate the effectiveness of the proposed method.
\end{itemize}

The organization of the paper is as follows:  \hyperref[sec:2]{Sec. 2} introduces basic concepts; \hyperref[sec:3]{Sec. 3} presents the quantum circuit partitioning method based on the QUBO model; \hyperref[sec:4]{Sec. 4} describes the transmission cost optimization method based on dynamic lookahead; \hyperref[sec:5]{Sec. 5} validates the effectiveness of the proposed method through experiments; \hyperref[sec:6]{Sec. 6} concludes the paper.

\section{\label{sec:2}Background}
In this section, the distributed architecture model and distributed quantum circuits are briefly introduced.

\subsection{Distributed Superconducting Architecture}
In recent years, the interconnect technology for superconducting qubits \cite{zhong2021deterministic,mirhosseini2020quantum,laracuente2022modeling,magnard2020microwave,qasymeh2022high,niu2023low} has made notable progress, enabling the interconnection of superconducting qubits across multiple quantum chips to expand the number of qubits within a single QPU. This interconnect technology typically employs superconducting coaxial cables \cite{zhong2021deterministic}, optical fibers \cite{mirhosseini2020quantum}, and microwave links \cite{laracuente2022modeling} to connect QPUs, facilitating the direct quantum state transmission in both low-temperature \cite{magnard2020microwave} and room-temperature \cite{qasymeh2022high} environments.

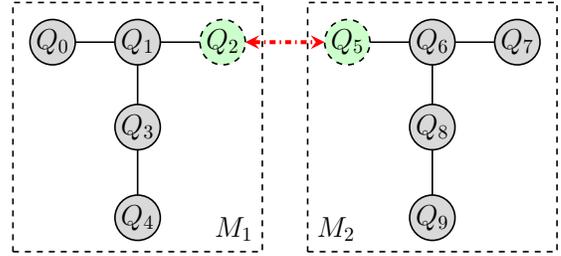
\begin{figure}[]%
	\centering
	\begin{adjustbox}{width=0.4\textwidth}
		\begin{tikzpicture}
			[>=stealth,node distance=1.9cm, thick, inner sep=0pt,minimum size=8mm,
			black/.style={circle,draw=black!100}]
			\node at (7.2,8.2) [] {\Large $M_1$};
			\node at (9,8.2) [] {\Large $M_2$};
			
			\draw [dashed] (3.3,7.8) rectangle (7.7,12.2);
			\node at (4,11.5) [circle,draw=black!100,fill=gray!30] {\Large $Q_0$};
			\node at (5.5,11.5) [circle,draw=black!100,fill=gray!30] {\Large$Q_1$};
			\node at (7,11.5) [circle,dashed,draw=black!100,fill=green!20] {\Large$Q_2$};
			\node at (5.5,10) [circle,draw=black!100,fill=gray!30] {\Large$Q_3$};
			\node at (5.5,8.4) [circle,draw=black!100,fill=gray!30] {\Large$Q_4$};
			\draw [thick] (5.5,10.4) -- (5.5,11.1);
			\draw [thick] (4.4,11.5) -- (5.1,11.5);
			\draw [thick] (5.9,11.5) -- (6.6,11.5);
			\draw [thick] (5.5,9.6) -- (5.5,8.8);

			\draw [dashed] (8.5,7.8) rectangle (12.9,12.2);
			\node at (9.2,11.5) [circle,dashed,draw=black!100,fill=green!20] {\Large$Q_5$};
			\node at (10.7,11.5) [circle,draw=black!100,fill=gray!30] {\Large$Q_6$};
			\node at (12.2,11.5) [circle,draw=black!100,fill=gray!30] {\Large$Q_7$};
			\node at (10.7,10) [circle,draw=black!100,fill=gray!30] {\Large$Q_8$};
			\node at (10.7,8.4) [circle,draw=black!100,draw=black!100,fill=gray!30] {\Large$Q_9$};

			\draw [thick] (10.7,10.4) -- (10.7,11.1);
			\draw [thick] (9.6,11.5) -- (10.3,11.5);
			\draw [thick] (11.1,11.5) -- (11.8,11.5);
			\draw [thick] (10.7,9.6) -- (10.7,8.8);			
			\draw [thick,dash dot,red,<->,line width=2pt] (7.4,11.5) -- (8.8,11.5);		
		\end{tikzpicture}
	\end{adjustbox}
 \caption{An example of a distributed quantum architecture which consists of two IBMQ\_quito architectures.}
 \label{fig1}
\end{figure}

Fig. \ref{fig1} provides an example of a distributed superconducting architecture model. \(M_1\) and \(M_2\) represent QPUs, interconnected through superconducting qubit interconnect technology, forming a distributed superconducting architecture. The dashed boxes contain the coupling graph for an individual QPU, with red dashed lines representing quantum channels. Dashed circles denote communication qubits \cite{cuomo2020towards}, used for transferring quantum states between different QPUs. The remaining qubits represent data qubits, utilized for storing and manipulating quantum states.

\subsection{Distributed Quantum Circuit}
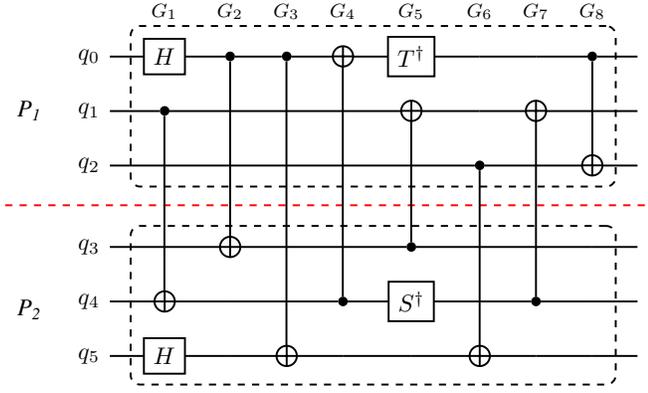
\begin{figure}[]
    \centering  
    \begin{adjustbox}{width=0.47\textwidth}
        \begin{tikzpicture}[>=stealth,baseline=0cm]
            \draw [dashed,thick,red] (0,0) -- (9.6,0);
        \end{tikzpicture}
        \hspace{-10cm}
        \begin{quantikz}[row sep={0.8cm,between origins},column sep=0.5cm]
            \lstick{$q_0$}  & \gate{H}\gategroup[3,steps=1,style={draw=none,rounded corners},background,label style={label position=left,anchor=north,xshift=-1.4cm}]{\textit{P\textsubscript{1}}}\gategroup[wires=1,steps=1,style={draw=none,rounded corners, inner sep=1pt}, label style={label position=above}]{\footnotesize $G_1$}\gategroup[3,steps=8,style={dashed,rounded corners,inner xsep=2pt,inner ysep=1pt},background]{} & \targ{}\gategroup[wires=1,steps=1,style={draw=none,rounded corners, inner sep=1pt}, label style={label position=above}]{\footnotesize $G_2$} & \ctrl{3}\gategroup[wires=1,steps=1,style={draw=none,rounded corners, inner sep=1pt}, label style={label position=above}]{\footnotesize $G_3$} & \gate{T^\dagger}\gategroup[wires=1,steps=1,style={draw=none,rounded corners, inner sep=1pt}, label style={label position=above}]{\footnotesize $G_4$} & \qw\gategroup[wires=1,steps=1,style={draw=none,rounded corners, inner sep=1pt}, label style={label position=above}]{\footnotesize $G_5$} & \targ{}\gategroup[wires=1,steps=1,style={draw=none,rounded corners, inner sep=1pt}, label style={label position=above}]{\footnotesize $G_6$} & \qw\gategroup[wires=1,steps=1,style={draw=none,rounded corners, inner sep=1pt}, label style={label position=above}]{\footnotesize $G_7$} & \targ{}\gategroup[wires=1,steps=1,style={draw=none,rounded corners, inner sep=1pt}, label style={label position=above}]{\footnotesize $G_8$} & \qw \\
            \lstick{$q_1$}  & \qw & \qw & \qw & \qw & \ctrl{4} & \qw & \qw & \qw& \qw\\
            \lstick{$q_2$}  & \targ{} & \qw & \qw& \ctrl{1} & \qw  & \qw & \ctrl{3} & \qw& \qw\\ [2.5ex]
            \lstick{$q_3$}  & \ctrl{-1}\gategroup[3,steps=1,style={draw=none,rounded corners},background,label style={label position=left,anchor=north,xshift=-1.4cm}]{\textit{P\textsubscript{2}}}\gategroup[3,steps=8,style={dashed,rounded corners,inner xsep=2pt,inner ysep=1pt},background]{} & \ctrl{-3} & \targ{} & \targ{} & \qw & \ctrl{-3} & \qw & \qw& \qw\\
            \lstick{$q_4$}  & \gate{H} & \qw & \qw & \qw & \qw & \gate{S^\dagger}  & \qw & \ctrl{-4}& \qw\\
            \lstick{$q_5$}  & \qw & \qw & \qw & \qw & \targ{} & \qw & \targ{} & \qw& \qw
        \end{quantikz}
    \end{adjustbox}
 \caption{An example of a distributed quantum circuit, where qubits \(q_0\) to \(q_5\) are divided into two partitions.}
 \label{fig2}
\end{figure}

Distributed quantum circuits are composed of several subcircuits with limited capacity, which are physically distanced from each other yet collaboratively simulate the functionality of large quantum circuits. Each subcircuit represents a partition, and each partition is mapped to a QPU. In executing distributed quantum circuits on a distributed architecture, two-qubit gates may act on qubits from different QPUs, termed as global gates. Those acting within the same QPU are called local gates. Local gates and single-qubit gates are executed directly within their respective partitions, while global gates require routing the quantum state to the same QPU for execution. Fig. \ref{fig2} shows an example of a distributed quantum circuit, where qubits \(q_0\) to \(q_5\) are divided into two partitions. Gates that span partitions are global gates, while those within the same partition are local gates.

\section{Quantum Circuit Partitioning Based on the QUBO Model}\label{sec:3}
The distributed quantum circuit partitioning problem involves dividing qubits into multiple partitions and minimizing the number of quantum gates acting between different partitions. This section transforms the problem into the minimum cut problem of a qubit-weighted graph, introduces a distributed quantum circuit partitioning method based on the QUBO model, and uses quantum optimization algorithms to solve this problem.

\subsection{Partitioning Cost Indicators}
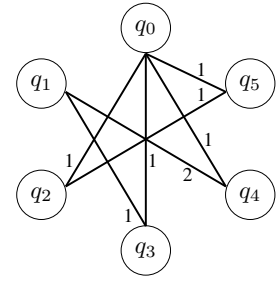
\begin{figure}%
    \centering
    \begin{adjustbox}{width=0.23\textwidth}
        \begin{tikzpicture}
            \node (q1) at (4,10) [circle,draw=black!100] {$q_1$};
            \node (q5) at (7,10) [circle,draw=black!100] {$q_5$};
            \node (q2) at (4,8.4) [circle,draw=black!100] {$q_2$};
            \node (q4) at (7,8.4) [circle,draw=black!100] {$q_4$};
            \node (q0) at (5.5,10.8) [circle,draw=black!100] {$q_0$};
            \node (q3) at (5.5,7.6) [circle,draw=black!100] {$q_3$};
            \draw [thick] (q0) -- (q3); 
            \draw [thick] (q5) -- (q2); 
            \draw [thick] (q2) -- (q3); 
            \draw [thick] (q0) -- (q4); 
            \draw [thick] (q1) -- (q5); 
            \draw (5.1,10.2) node {\footnotesize{1}};
            \draw (6.3,9.2) node {\footnotesize{1}};  
            \draw (4.85,8.2) node {\footnotesize{2}}; 
            \draw (5.7,8.9) node {\footnotesize{3}}; 
            \draw (4.89,9.1) node {\footnotesize{1}}; 
        \end{tikzpicture}
    \end{adjustbox}
         \caption{Qubit-weighted graph with nodes as qubits, edges as quantum gates acting between qubits, and weights as the number of quantum gates.}
     \label{fig3}
    \end{figure}

The distributed quantum circuit partitioning problem can be converted into the minimum cut problem of a qubit-weighted graph. Fig. \ref{fig3} shows the qubit-weighted graph \(G(V, E, W)\) for the distributed quantum circuit depicted in Fig. \ref{fig2}, where the set of vertices \(V\) represents logical qubits, the set of edges \(E\) represents two-qubit quantum gates acting between pairs of logical qubits, and the set of weights \(W\) represents the number of these quantum gates. Partitioning the qubit-weighted graph and dividing the vertices, i.e., qubits, into \(K\) subsets, achieves an equivalent partition of the distributed quantum circuit. The indicator for partitioning distributed quantum circuits typically aims to minimize global gates or transmission cost. Since the transmission cost is closely related to the number of global gates, this work chose to use the number of global gates as an important indicator of circuit partitioning. In the qubit-weighted graph, this is represented by minimizing the edge weight between sets of vertices, which constitutes a graph minimum cut problem.

To precisely quantify transmission cost, the following definition is introduced.

\textbf{Definition 1:} In distributed quantum computing, transmission cost refers to the resource consumption, time delay, and error control overhead incurred when transmitting quantum states between different quantum computing nodes. In a simplified model, the transmission cost $TC$ can be expressed as the number of quantum state transmissions:
\begin{equation}
TC =  c \cdot T
\label{eq:transmission_cost}
\end{equation}
where $T$ represents the number of quantum state transmissions, and $c$ represents the unit cost of each quantum state transmission. 

In this definition, it is assumed that the cost of each quantum state transmission is constant, making the transmission cost a linear function of the number of quantum state transmissions. To further simplify, if $c$ is set to 1, then $TC$ equals the number of quantum state transmissions.



To adhere to the no-cloning theorem \cite{jozsa2004illustrating} and ensure the integrity of quantum information, when a quantum state of a global gate is transferred from the original QPU to the target QPU, it must be re-transferred back to the original QPU to avoid affecting the execution of subsequent gates in the original QPU. Therefore, each global gate requires two transmissions, making the upper bound of the transmission cost equal to twice the number of global gates. However, once a quantum state is transferred from the original partition to the target partition, it can continue to be used by subsequent quantum gates in the target partition. This model, known as the merged transfer model, offers a lower transmission cost than twice the number of global gates as the transmission cost. Therefore, it is widely used in optimizing transmission cost in  \cite{ghodsollahee2021connectivity,wu2022autocomm} and \cite{cheng2023optimization}. The application of the merged transfer model originates from a fundamental observation in distributed quantum circuits, where substantial communication is concentrated between specific pairs of nodes or qubits. This phenomenon is analogous to the concept of burst communication in classical computing. Burst communication mechanisms in distributed quantum computing are thoroughly discussed in \cite{wu2022autocomm}. To further enhance the optimization of the transmission cost, the effect of merged transfer is integrated into the cost indicator for distributed quantum circuit partitioning. Since the optimized transmission cost cannot be directly represented by a specific function, a global gate dispersion relation graph is constructed to evaluate the indirect relationship between the number of global gates and the transmission cost.

\textbf{Definition 2:} In the qubit-weighted graph \(G(V, E, W)\), after all qubits \(V\) are partitioned, edges within the same partition \(E_{lg}\) are removed, retaining only the edges between partitions \(E_{gg}\). The updated edges \(E\) in the qubit-weighted graph represent global gates. This graph is referred to as the global gate dispersion relation graph, denoted as \(G'(V, E_{gg}, W_{g g})\).

In the global gate dispersion relation graph \(G'(V, E_{gg}, W_{g g})\), a global gate dispersion function is defined to assess the impact of the number and dispersion of global gates on the transmission cost, as shown in Eq. \eqref{1}.
\begin{equation}
  F g g=\frac{\sum W_{g g}}{N_{E_{g g}}}
 \label{1}
\end{equation}
Here, \(\sum W_{g g}\) denotes the total number of global gates, while \(N_{E_{gg}}\) represents the number of edges between partitions, indicating the variety of qubit pairs affected by global gates. A higher \(F_{gg}\) indicates that, on average, more global gates are associated with each edge between partitions, suggesting a denser distribution of global gates. In addition, a higher \(F_{gg}\) also implies that fewer qubits are involved in the partition interaction. This dense distribution of global gates among a smaller number of sub-qubits is more compatible with the merged transfer model, potentially reducing the overall transmission cost. Specifically, when global gates are concentrated on certain edges and executed consecutively, the transmission operations on these edges can be optimized through merged transfers, thereby decreasing the total transmission cost.
\begin{figure}[]
    \centering
    \setcounter{subfigure}{0}
    \subfigure[]{ 
        \begin{adjustbox}{width=0.22\textwidth} 
        \begin{tikzpicture}
            
            \draw[dashed] (0,0) ellipse (3cm and 0.8cm);
            
            \draw[] (0,0) circle (0.4cm);
            \node[anchor=center] at (0,0) {\huge $q_1$};
            
            \draw[] (2,0) circle (0.4cm);
            \node[anchor=center] at (2,0) {\huge$q_2$};
            
            \draw[] (-2,0) circle (0.4cm);
            \node[anchor=center] at (-2,0) {\huge$q_0$};

            \draw[dashed] (0,-3) ellipse (3cm and 0.8cm);
            
            \draw[] (0,-3) circle (0.4cm);
            \node[anchor=center] at (0,-3) {\huge$q_4$};
            
            \draw[] (2,-3) circle (0.4cm);
            \node[anchor=center] at (2,-3) {\huge$q_5$};
            
            \draw[] (-2,-3) circle (0.4cm);
            \node[anchor=center] at (-2,-3) {\huge$q_3$};

            \draw (-2,-0.4) -- (-2,-2.6);
            \node at(-2.3,-1.5) {\large$3$};

            \draw (0,-0.4) -- (2,-2.6);
            \node at(1.52,-1.7) {\large$1$};
            
            \draw (-2,-0.4) -- (0,-2.6);
            \node at(-1,-1.2) {\large$1$};
            
            \draw (2,-0.4) -- (2,-2.6);
            \node at(2.2,-1.5) {\large$1$};
            
            \draw (-2,-2.6) -- (2,-0.4);
            \node at(0.3,-1.6) {\large$2$};
        \end{tikzpicture}
        \end{adjustbox}
         \label{fig4a}
    } 
    \subfigure[]{  
        \begin{adjustbox}{width=0.22\textwidth}
         \begin{tikzpicture}
     
     \draw[dashed] (0,0) ellipse (3cm and 0.8cm);
     
     \draw[] (0,0) circle (0.4cm);
     \node[anchor=center] at (0,0) {\huge$q_1$};
     
     \draw[] (2,0) circle (0.4cm);
     \node[anchor=center] at (2,0) {\huge$q_4$};
     
     \draw[] (-2,0) circle (0.4cm);
     \node[anchor=center] at (-2,0) {\huge$q_0$};

     \draw[dashed] (0,-3) ellipse (3cm and 0.8cm);
     
     \draw[] (0,-3) circle (0.4cm);
     \node[anchor=center] at (0,-3) {\huge$q_3$};
     
     \draw[] (2,-3) circle (0.4cm);
     \node[anchor=center] at (2,-3) {\huge$q_5$};
     
     \draw[] (-2,-3) circle (0.4cm);
     \node[anchor=center] at (-2,-3) {\huge$q_2$};

     \draw (-2,-0.4) -- (0,-2.6);
     \node at(-1.52,-1.4) {\large$3$};

     \draw (0,-0.4) -- (2,-2.6);
     \node at(0.62,-1.4) {\large$1$};

 \end{tikzpicture}
    \end{adjustbox}
\label{fig4b}
	}
 \caption{Global gate dispersion relation graph. The values of (a) and (b) for the global gate dispersion function \(F_{gg}\) are 8/5 and 4/2, respectively.}
 \label{fig4}

\end{figure}
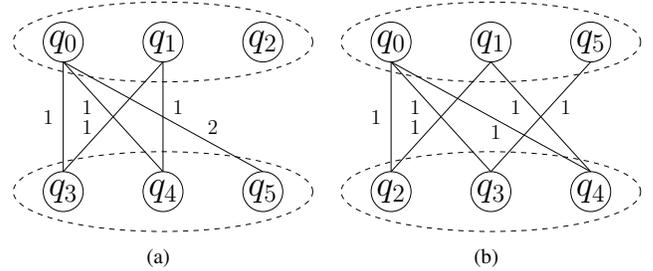

Taking the global gate dispersion relation graph in Fig. \ref{fig4} as an example to illustrate the impact of dispersion on the transmission cost. Fig. \ref{fig4}(a) shows the scenario in the qubit-weighted graph from Fig. \ref{fig3} where qubits \(q_0\), \(q_1\), and \(q_2\) are partitioned into one section, and \(q_3\), \(q_4\), and \(q_5\) into another. In this case, there are a total of 8 global gates, 5 edges,
thus, the global gate dispersion function \(F_{gg} = \frac{8}{5}\), implying that the edges of each partition are associated with 1.6 global gates on average. The final transmission cost under this partitioning is 6. Similarly, in Fig. \ref{fig4}(b), the transmission cost is 2, with the global gate dispersion function \(F_{gg} = \frac{4}{2}\). Therefore, a larger global gate dispersion function may result in a lower transmission cost.

Therefore, after considering both the number of global gates and the global gate dispersion function, the cost indicator for distributed quantum circuit partitioning is a comprehensive indicator. It optimizes both the number of global gates and their distribution, which helps to reduce transmission costs.

\subsection{Optimization of Quantum Circuit Partitioning Based on the QUBO Model}
The minimum cut problem in graph theory, aimed at identifying the smallest edge set disconnecting specified vertex sets, presents a combinatorial optimization challenge. Though algorithms like Fiduccia-Mattheyses algorithm and Kernighan-Lin algorithm as well as clustering methods such as K-means and Spectral Clustering offer solutions, they may not always achieve the global optimum due to the problem’s NP-hard nature, requiring exponential solution time. To tackle this challenge, this work maps the problem onto the QUBO model \cite{dunning2018works}. The QUBO model, utilizing the quadratic form of binary variables along with appropriate constraints and objective functions, employs quantum annealing algorithms to accelerate the solution process.

Specifically, for a given qubit-weighted graph, a set of binary variables \(X_{ik}\) is defined, as shown in Eq. \eqref{3},
\begin{equation}
	X_{ik} =
	\begin{cases}
		1, &  q_i \in P_k \\
		0, &  q_i \notin P_k
	\end{cases}
 \label{3}
\end{equation}
where $i \in \{0,1,\dots,N-1\},  k \in \{0,1,\dots,K-1\}$, with \(N\) representing the number of qubits and \(K\) representing the number of partitions. Each variable \(X_{ik}\) represents whether a vertex \(X_i\) belongs to a set \(P_k\), meaning whether the qubit \(q_i\) is partitioned to the \(k\)-th partition. When \(X_{ik} = 1\), it indicates that the qubit \(q_i\) is assigned to partition \(P_k\).

\begin{table}[]
\centering
\caption{Truth table when vertices $i$ and $j$ are assigned to the same partition and different partitions.}
 \begin{minipage}{0.2\textwidth}
        \footnotesize
        \centering
        \begin{tabular}{c|ccc}  
            \toprule
           \cellcolor[HTML]{E5E5E5} \textit{k} & \cellcolor[HTML]{E5E5E5}\textit{X\textsubscript{i}} & \cellcolor[HTML]{E5E5E5}\textit{X\textsubscript{j}} & \cellcolor[HTML]{E5E5E5}\textit{X\textsubscript{i}X\textsubscript{j}} \\
            \midrule
            0       & 0   & 0   & 0   \\
            1       & 0   & 0   & 0   \\
            2       & 0   & 0   & 0   \\
            \vdots & \vdots & \vdots & \vdots \\
            \textit{k}  & 1   & 1   & 1   \\
            \vdots & \vdots & \vdots & \vdots \\
            \textit{K-1} & 0   & 0   & 0   \\
            \bottomrule
        \end{tabular}
    \end{minipage}
       \begin{minipage}{0.2\textwidth}
        \centering
        \renewcommand{\arraystretch}{0.8}
        \begin{tabular}{c|ccc}  
            \toprule
            \cellcolor[HTML]{E5E5E5}\textit{k} & \cellcolor[HTML]{E5E5E5}\textit{X\textsubscript{i}} & \cellcolor[HTML]{E5E5E5}\textit{X\textsubscript{j}} & \cellcolor[HTML]{E5E5E5}\textit{X\textsubscript{i}X\textsubscript{j}} \\
            \midrule
            0       & 0   & 0   & 0   \\
            \vdots & \vdots & \vdots & \vdots \\
            \textit{k\textsubscript{1}}  & 0   & 1   & 0   \\
            \vdots & \vdots & \vdots & \vdots \\
            \textit{k\textsubscript{2}}  & 1   & 0   & 0  \\
            \vdots & \vdots & \vdots & \vdots \\
            \textit{K-1} & 0   & 0   & 0   \\
            \bottomrule
        \end{tabular}
    \end{minipage}
   
\label{xik truth table}
\end{table}

To minimize the number of global gates, i.e., to minimize the sum of weights between partitions in the qubit-weighted graph, it involves assessing whether the two qubits \(X_i\) and \(X_j\) acting on each edge \(E_{ij}\) in the qubit-weighted graph are in different partitions. As can be seen from Table~\ref{xik truth table}, under the binary variable \(X_{ik}\), qubits \(X_i\) and \(X_j\) are in the same partition $k$ if and only if \(\sum_{k=0}^{K-1} X_{ik}  X_{jk} = 1\), qubits \(X_i\) and \(X_j\) are in different partitions if and only if \(\sum_{k=0}^{K-1} X_{ik}  X_{jk} = 0\). Therefore, \(1\!-\!\sum_{k=0}^{K-1} X_{ik} X_{jk}\) can determine if qubits \(X_i\) and \(X_j\) are in different partitions. \(X_i\) and \(X_j\) are in different partitions when \(1\!-\! \sum_{k=0}^{K-1} X_{ik}  X_{jk} = 1\), i.e., the quantum gates acting on \(X_i\) and \(X_j\) are global gates. The number of global gates after partitioning is denoted as $F_1$, as shown in Eq. \eqref{4},
\begin{equation}
    F_1=\sum W_{gg}={\sum_{i,j\in E_{ij}}W_{ij}(1-\sum_{k=0}^{K-1}X_{ik}X_{jk})} 
\label{4}
\end{equation}
where \(i\) and \(j\) denote the indices of any two qubits \(q_i\) and \(q_j\), \(W_{ij}\) represents the number of quantum gates acting on qubits \(q_i\) and \(q_j\), $W_{ij}(1-\sum_{k=0}^{K-1}X_{ik}X_{jk})$ represents the number of global gates acting on qubits \(q_i\) and \(q_j\). A lower value of $F_1$ will result in lower transmission cost.

The global gate dispersion function $F_{gg}$ can likewise be described using the binary variable $X_{ik}$ in a qubit-weighted graph. The function describing $F_{gg}$ in terms of the binary variable $X_{ik}$ is denoted as $F_2$, while $F_2$ is expected to be as large as possible. The transformed objective function $F_2$ for the global gate dispersion function is shown in Eq. \eqref{5}.
\begin{equation}
    \begin{aligned}
 F_2 & = \frac{\sum W_{g g}}{N_{E_{g g}}}=\frac{\displaystyle\sum_{i,j\in E_{ij}}W_{ij}(1-\sum_{k=0}^{K-1}X_{ik}X_{jk})}{\displaystyle\sum_{i,j\in E_{ij}}(1-\sum_{k=0}^{K-1}X_{ik}X_{jk})} \\
\end{aligned}
\label{5}
\end{equation}

Since the QUBO model cannot solve the objective function in division form, a transformation of $F_2$ in division form is required. Multiple iterative transformations are used in \cite{matsumoto2022distance} to convert the objective function from a complex division form to one suitable for subtraction, but this method requires multiple optimizations and adjustments. In order to simplify the calculation process, this work transform $F_2$ into a subtraction form $\hat{F_2}$, shown as in Eq. \eqref{51}.
\begin{equation}
    \begin{aligned}
 \hat{F_2} & =\sum_{i,j\in E_{ij}}W_{ij}(1-\sum_{k=0}^{K-1}X_{ik}X_{jk})  \\
&  -\sum_{i,j\in E_{ij}}(1-\sum_{k=0}^{K-1}X_{ik}X_{jk})
\end{aligned}
\label{51}
\end{equation}

This transformation can improve the computational efficiency and satisfy the need to quickly find an approximate solution rather than pursuing an exact solution. To further validate the relationship between $F_2$ and $\hat{F_2}$, numerical simulations are conducted for different configurations. The results show a positive correlation between the two functions, especially as the number of qubits and gates increases. This confirms that the transformation from $F_2$ to $\hat{F_2}$ preserves the original optimization goal. 

Further, consider the constraints. The constraints for the partitioning of distributed quantum circuits are that each qubit can only be assigned to one partition, and the number of qubits within each partition should vary as little as possible to ensure load balancing across the distributed system. Therefore, the constraint conditions for the partitioning of distributed quantum circuits can be represented by Eq. \eqref{6}, 
\begin{equation}
    \text{s.t.} 
\begin{cases}
\displaystyle\sum_{k=0}^{K-1} X_{i k}=1, \forall i \in \{0,1,\dots,N\!-\!1\} \\
\displaystyle \left\lfloor\frac{N}{K}\right\rfloor\!-\!\rho \leq\sum_{i=0}^{N-1} X_{ik} \leq  \left\lceil\frac{N}{K}\right\rceil\!+\!\rho, \forall k \in \{0,1,\dots,K\!-\!1\} 
\end{cases}
\label{6}
\end{equation}
where the constraints in the first part ensure that any qubit \(q_i\) can only be assigned to one partition; moreover, \(N/K\) represents the average size of a partition; \(\rho\) represents the load balance tolerance, which controls the variance in the number of qubits across different partitions; and the constraints in the second part ensure that the size of each partition after division does not differ from the average \(N/K\) by more than the load balancing tolerance \(\rho\).

In QUBO models, directly enforcing such linear equality constraints poses challenges, as QUBO models require the objective function to be quadratic and free of explicit constraints. To address this, the linear equality constraint in the first part of Eq. \eqref{6} is transformed into a quadratic penalty term, resulting in the following objective function:
\begin{equation}
    st_1= \sum_{i=0}^{N-1}\left(\sum_{k=0}^{K-1}X_{ik}-1 \right)^2
    \label{st1}
\end{equation}
This transformation transforms the constraints into a squared difference penalty term for each $i$ and sums all these penalty terms to form a new objective function. When the objective function is minimized, the penalty term is zero only if $\sum_{k=0}^{K-1} X_{i k}$ corresponding to each $i$ is equal to 1, thus ensuring that the original constraint is satisfied.

In order to transform the inequality constraints on $ \sum_{i=0}^{N-1} X_{ik} \leq  \left\lceil\frac{N}{K}\right\rceil+\rho$ in the second part of the original constraints in Eq. \eqref{6} into constraints in the form of equations suitable for the QUBO, this work introduces slack variables \( S_t \in \{0,1\} \) and represents the constraint as Eq. \eqref{st2},
\begin{equation}
     st_{2\leq}= \sum_{k=0}^{K-1}\left( \sum_{i=0}^{N-1} X_{ik} + \sum_{m=0}^{\left\lceil \log_2 \left( \left\lceil\frac{N}{K}\right\rceil + \rho + 1 \right) \right\rceil - 1} 2^m Y_m - (\left\lceil\frac{N}{K}\right\rceil + \rho) \right)^2 
    \label{st2}
\end{equation}
where the set of $Y\subseteq S$. Specifically, the slack variables \( Y_m \) are used to encode the upper bound of the constraint in binary form, and the number of \( Y_m \) is $\lceil \log_2 \left(\left\lceil\frac{N}{K}\right\rceil + \rho + 1 \right)\rceil$. In this way, by choosing the appropriate slack variables, it is always possible to make the values of \(\sum_{i=0}^{N-1} X_{ik}\) satisfy the constraints of $\leq$. Similarly, the constraints on $ \sum_{i=0}^{N-1} X_{ik} \geq\left\lfloor\frac{N}{K}\right\rfloor-\rho$ are handled in the same way, and denoted as $st_{2\geq}$.

After defining the two objective functions \( F_1 \) and \( \hat{F_2} \) along with constraints, the next step is to integrate these components into a unified overall optimization objective function. By introducing appropriate weighting coefficients, the objective functions and penalty terms are combined in a manner that ensures the effective reduction of the number of global gates and the increase of global gate dispersion, while simultaneously enforcing the load balancing requirements for partitioning. The overall QUBO objective function is given by Eq. \eqref{7},
\begin{equation} 
    \begin{aligned} 
    F &= F_1 - \varphi \hat{F_2} + \lambda_1 st_1+ \lambda_2 st_{2\leq} + \lambda_2 st_{2\geq}\\
    &= (1-\varphi)\sum_{i,j\in E_{ij}}W_{ij}(1-\sum_{k=0}^{K-1}X_{ik}X_{jk}) \\
    &-\varphi\sum_{i,j\in E_{ij}}(1-\sum_{k=0}^{K-1}X_{ik}X_{jk})\\
    &+ \lambda_1 \sum_{i=0}^{N-1}\left(\sum_{k=0}^{K-1}X_{ik}-1 \right)^2 \\
    &+ \lambda_2 \sum_{k=0}^{K-1}\left(\sum_{i=0}^{N-1}X_{ik}+\!\sum_{m=0}^{\left\lceil \log_2 \left(\left\lceil\frac{N}{K}\right\rceil\!+\!\rho\!+\!1 \right) \right\rceil\!-\!1} 2^mY_m -(\left\lceil\frac{N}{K}\right\rceil\!+\!\rho)\right)^2\\
    &+ \lambda_2 \sum_{k=0}^{K-1}\left(\sum_{i=0}^{N-1}X_{ik}\!-\!\sum_{n=0}^{\left\lceil \log_2 \left(\left\lfloor\frac{N}{K}\right\rfloor\!+\!  \rho\!+\!1 \right) \right\rceil\!-\!1} 2^nZ_n -(\left\lfloor\frac{N}{K}\right\rfloor\!-\!\rho)\right)^2
    \end{aligned}
    \label{7}
\end{equation}
where $\varphi$ is used to control the weights of the two objective functions to weight the two optimization objectives. The negative sign before $\varphi$ ensures that the optimization objective of $\hat F_2$ aligns with the overall minimization of $F$. By adding penalty coefficients \(\lambda_1\) and \(\lambda_2\) to the constraint conditions, so that, as far as possible, the final circuit partitioning result satisfies the constraints.

To transform the overall optimization objective function \( F \) into the QUBO model's quadratic form \( \mathbf{x}^T Q \mathbf{x} \), \( F \) is expanded to include only quadratic terms, ensuring each term involves at most two binary variables. Linear terms in \( F \) are converted into quadratic terms using the binary variable property \( X_{ik}^2 = X_{ik} \). Then, each binary variable \( X_{ik} \) is assigned a unique index, and the coefficients of the quadratic terms are systematically mapped to their respective positions in the QUBO matrix \( Q \). This methodical mapping results in a QUBO matrix \( Q \) that fully represents the original optimization problem. After being transformed into a QUBO matrix, the objective function in Eq. \eqref{7} can be deployed for execution on an annealing machine. The QUBO model has high solution efficiency, especially when combined with quantum annealing algorithms, which can accelerate the solution speed.

\section{Optimization of Transmission Cost Based on Dynamic Lookahead}\label{sec:4}
This section introduces a dynamic lookahead method for selecting transmission qubits for global gates, distinguished by three types of impacts that the selection of transmission qubit has on subsequent quantum gates. This method determines the lookahead window size for the current global gate based on the relationship between subsequent gates and the transmission qubits. It then constructs an impact cost function to evaluate the transmission qubits, thereby optimizing the transmission cost of distributed quantum circuits.
\subsection{Transmission Qubits of Global Gates}
In distributed quantum circuits, since global gates operate across different QPUs, it is necessary to transfer the quantum state of one qubit of the global gate to the partition where another qubit resides, in order to reconstruct it into a local gate for execution.

\textbf{Definition 3:} To execute a global gate in distributed quantum circuits, the quantum state of the qubit acting on the global gate must be transferred to the partition where the other qubit is located. At this point, the qubit being transferred is referred to as the transmission qubit.

\begin{figure}[]
	\centering
	\subfigure[]{  
		\begin{adjustbox}{width=0.22\textwidth}
			\begin{tikz}
				\draw [dashed,thick,red] (-3,1) -- (1.45,1);
			\end{tikz}
			\hspace{-4.9cm}
			\begin{quantikz}[row sep={0.6cm,between origins}]
				\lstick{$q_{i_1}$}  & \ctrl{6}\gategroup[4,steps=3,style={draw=none,rounded corners},background,label style={label position=left,anchor=north,xshift=-1.5cm}]{\textit{P\textsubscript{I}}}\gategroup[wires=1,steps=1,style={draw=none,rounded corners, inner sep=1pt}, label style={label position=above}]{\footnotesize $G_1$}  & \ctrl{7}\gategroup[wires=1,steps=1,style={draw=none,rounded corners, inner sep=1pt}, label style={label position=above}]{\footnotesize $G_2$}   & \ctrl{6}\gategroup[wires=1,steps=1,style={draw=none,rounded corners, inner sep=1pt}, label style={label position=above}]{\footnotesize $G_3$}   & \qw\\
				\lstick{$\vdots$}  & \wave&&& \rstick{$\vdots$}\\
				\lstick{$q_{i_{n-1}}$}  & \qw & \qw  & \qw  & \qw \\
				\lstick{$q_{i_n}$}  & \qw & \qw  & \qw  & \qw \\
				\lstick{$q_{j_1}$}  & \qw\gategroup[4,steps=3,style={draw=none,rounded corners},background,label style={label position=left,anchor=north,xshift=-1.5cm}]{\textit{P\textsubscript{J}}}  & \qw  & \qw  & \qw \\
				\lstick{$\vdots$}  & \wave&&& \rstick{$\vdots$}\\
				\lstick{$q_{j_{n-1}}$}  & \targ{} & \qw  & \targ{}  & \qw \\
				\lstick{$q_{j_n}$}  & \qw & \targ{} & \qw  & \qw
			\end{quantikz}
   \label{33a}
		\end{adjustbox}
	}
	\subfigure[]{  
		\begin{adjustbox}{width=0.22\textwidth}
			\begin{tikz}
				\draw [dashed,thick,red] (-3,1) -- (1.45,1);
			\end{tikz}
			\hspace{-4.9cm}
			\begin{quantikz}[row sep={0.6cm,between origins}]
				\lstick{$q_{i_1}$}  & \ctrl{4}\gategroup[4,steps=3,style={draw=none,rounded corners},background,label style={label position=left,anchor=north,xshift=-1.5cm}]{\textit{P\textsubscript{I}}}\gategroup[wires=1,steps=1,style={draw=none,rounded corners, inner sep=1pt}, label style={label position=above}]{\footnotesize $G_1$}  & \qw\gategroup[wires=1,steps=1,style={draw=none,rounded corners, inner sep=1pt}, label style={label position=above}]{\footnotesize $G_2$}   & \qw\gategroup[wires=1,steps=1,style={draw=none,rounded corners, inner sep=1pt}, label style={label position=above}]{\footnotesize $G_3$}& \qw\\
				\lstick{$\vdots$}  & \wave&&& \rstick{$\vdots$}\\
				\lstick{$q_{i_{n-1}}$}  & \qw & \qw  & \ctrl{5}  & \qw\\
				\lstick{$q_{i_n}$}  & \qw & \ctrl{3}  & \qw  & \qw\\
				\lstick{$q_{j_1}$}  & \targ{}\gategroup[4,steps=3,style={draw=none,rounded corners},background,label style={label position=left,anchor=north,xshift=-1.5cm}]{\textit{P\textsubscript{J}}}  & \qw  & \qw  & \qw\\
				\lstick{$\vdots$}  & \wave&&& \rstick{$\vdots$}\\
				\lstick{$q_{j_{n-1}}$}  & \qw & \targ{}  & \qw  & \qw\\
				\lstick{$q_{j_n}$}  & \qw & \qw  & \targ{}  & \qw
			\end{quantikz}
   \label{33b}
		\end{adjustbox}
	}      
 \caption{Scenarios for using the merged transfer model. (a) If selecting \(q_{i_1}\) as the transmission qubit, it matches the merged transfer model. (b) Unable to match the merged transfer model.}
 \label{333}

\end{figure}
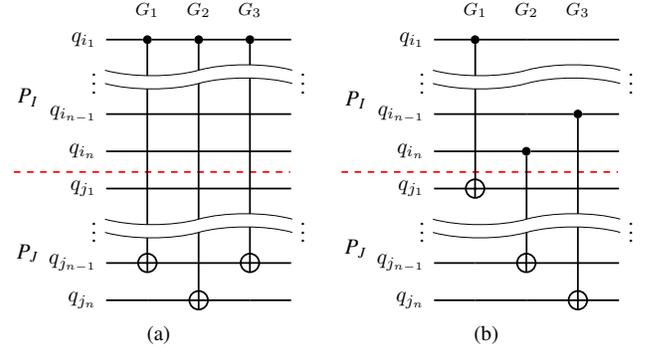

To reduce the transmission cost in distributed quantum circuits, a merged transfer model is summarized in \cite{cheng2023optimization}. In this model, global gates that share the same qubits and are adjacent can be restructured into local gates and executed through a single instance of state transmission. This model effectively reduces transmission cost and significantly enhances the efficiency of distributed quantum computing. As shown in Fig. \ref{333}(a), the global gates \(G_1\), \(G_2\), and \(G_3\) act on the same qubit \(q_{i_1}\), and the partitions of the other qubits involved in these gates are also the same. By selecting qubit \(q_{i_1}\) as the transmission qubit, all can be converted into local gates using merged transfer model. In contrast, as shown in Fig. \ref{333}(b), the global gates \(G_1\), \(G_2\), and \(G_3\) do not act on the same qubit, thus requiring sequential transmission for each global gate to convert them into local gates.

Currently, there are two prevalent methods for optimizing the transmission cost of distributed quantum circuits using the merged transfer model: meta-heuristic methods \cite{houshmand2020evolutionary,chen2023routing} and greedy strategies \cite{ghodsollahee2021connectivity,cheng2023optimization}. Meta-heuristic methods utilize intelligent approaches to dynamically adjust the transmission qubits for each gate to match the merged transfer model. This method may not converge to the optimal solution and can be unstable. Greedy strategies, targeting the current best solution, progressively traverse adjacent gates in the circuit from left to right, selecting the transmission qubits by assessing if adjacent or nearest gates act on the same qubits.

\begin{figure}
		\centering  
		\begin{adjustbox}{width=0.36\textwidth}
			\begin{tikzpicture}[>=stealth,baseline=0cm]
				\draw [dashed,thick,red] (-0.1,-0.2) -- (5.9,-0.2);
			\end{tikzpicture}
			\hspace{-6.4cm}
			\begin{quantikz}[row sep={0.6cm,between origins},column sep=0.5cm]
				\lstick{$q_0$}  & \ctrl{3}\gategroup[3,steps=1,style={draw=none,rounded corners},background,label style={label position=left,anchor=north,xshift=-1.4cm}]{\textit{P\textsubscript{1}}}\gategroup[wires=1,steps=1,style={draw=none,rounded corners, inner sep=1pt}, label style={label position=above}]{\footnotesize $G_1$}\gategroup[1,steps=2,style={dashed,rounded corners,fill=blue!40, inner xsep=2pt, inner ysep=-0.08cm, fill opacity=0.5},background,label style={label position=below,anchor=north,xshift=0.55cm,yshift=0.1cm}]{} & \ctrl{4}\gategroup[wires=1,steps=1,style={draw=none,rounded corners, inner sep=1pt}, label style={label position=above}]{\footnotesize $G_2$} & \qw\gategroup[wires=1,steps=1,style={draw=none,rounded corners, inner sep=1pt}, label style={label position=above}]{\footnotesize $G_3$} & \qw\gategroup[wires=1,steps=1,style={draw=none,rounded corners, inner sep=1pt}, label style={label position=above}]{\footnotesize $G_4$} & \qw\gategroup[wires=1,steps=1,style={draw=none,rounded corners, inner sep=1pt}, label style={label position=above}]{\footnotesize $G_5$} & \qw\\
				\lstick{$q_1$}  & \qw & \qw & \qw & \targ{} & \ctrl{3} & \qw \\
				\lstick{$q_2$}  & \qw & \qw & \targ{}& \qw & \qw  & \qw\\
				\lstick{$q_3$}  & \targ{}\gategroup[2,steps=1,style={draw=none,rounded corners},background,label style={label position=left,anchor=north,xshift=-1.4cm}]{\textit{P\textsubscript{2}}}\gategroup[1,steps=4,style={dashed,rounded corners,fill=green!40, inner xsep=2pt, inner ysep=-0.08cm, fill opacity=0.5},background,label style={label position=below,anchor=north,xshift=0.55cm,yshift=0.1cm}]{} & \qw & \ctrl{-1} & \ctrl{-2} & \qw & \qw \\
				\lstick{$q_4$}  & \qw & \targ{} & \qw & \qw & \targ{} & \qw 
			\end{quantikz}
		\end{adjustbox}
        \caption{Example of the impact of transmission qubits on transmission cost. The final transmission cost of using \(q_0\) as the transmission qubit is 6, while the final transmission cost of using \(q_3\) is 4.}
        \label{A}
\end{figure}
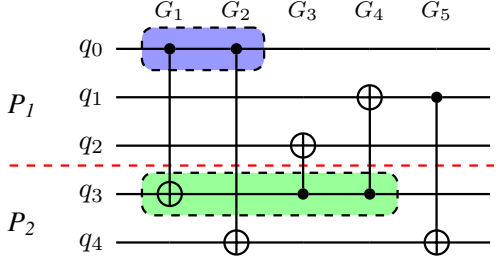

The greedy strategy sequentially traverses gates from \(G_1\) onwards, selecting appropriate qubits for transmission, as shown in Fig. \ref{A}. Using \(q_0\) as the transmission qubit for \(G_1\), \(G_1\) and \(G_2\) are merged transferred, then \(q_3\) meets the requirements for merged transferring \(G_3\) and \(G_4\), while \(q_1\) or \(q_4\) is used for executing \(G_5\), resulting in a transmission cost of 6. If \(q_3\) is used as the transmission qubit for \(G_1\), \(G_1\), \(G_3\), and \(G_4\) can be converted into local gates, and then using \(q_4\) as the transmission qubit for \(G_2\) and \(G_5\) reduces the transmission cost to 4. This demonstrates that judicious selection of transmission qubits can reduce transmission cost.

\subsection{Strategy for Selecting Transmission Qubits}
The selection of transmission qubits for global gates affects the transmission cost of subsequent quantum gates, thereby influencing the overall transmission cost of distributed quantum circuits. Selecting optimal transmission qubits can reduce transmission cost in the circuit, making it necessary to differentiate the impact of transmission qubits on subsequent quantum gates.

\textbf{Definition 4:} In distributed quantum circuits, the choice of different transmission qubits for executing a global gate will result in transmission cost variations for subsequent quantum gates. This variation is defined as the impact factor, denoted by \(E_q\).

The impact factor \(E_q\) can take on three different values, each corresponding to a distinct scenario. The values of the three impact factors \(E_q\) for the global gate are as shown in Eq. \eqref{8}.
\begin{equation}
    E_q = 
\begin{cases} 
1 & \text{positive impact} \\
0 & \text{no impact} \\
-1 & \text{negative impact}
\end{cases}
\label{8}
\end{equation}

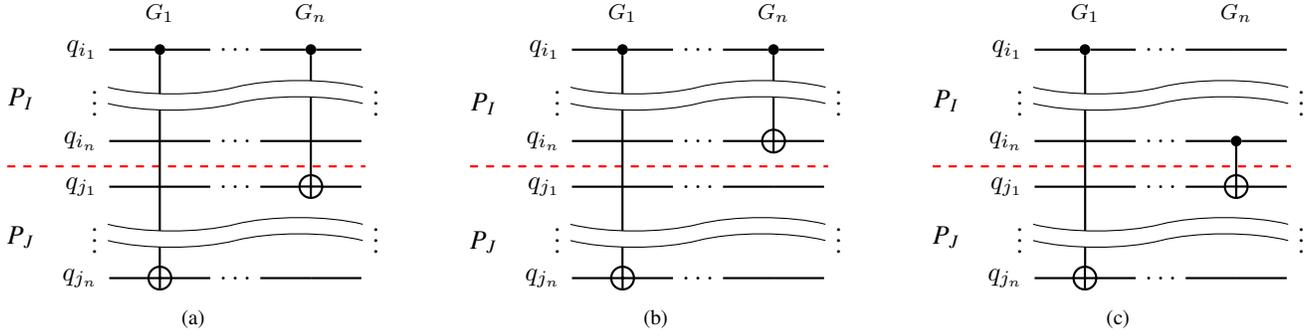
\begin{figure*}[]
	\centering
	\setcounter{subfigure}{0}
	\subfigure[]{  
			\begin{adjustbox}{width=0.28\textwidth}
		\begin{tikz}
			\draw [dashed,thick,red] (-3,1) -- (1.7,1);
		\end{tikz}
		\hspace{-5.2cm}
		\begin{quantikz}[row sep={0.6cm,between origins}]
			\lstick{$q_{i_1}$}  & \ctrl{5}\gategroup[3,steps=3,style={draw=none,rounded corners},background,label style={label position=left,anchor=north,xshift=-1.4cm}]{\textit{P\textsubscript{I}}}\gategroup[wires=1,steps=1,style={draw=none,rounded corners, inner sep=1pt}, label style={label position=above}]{\footnotesize $G_1$} &\ \ldots\ \qw & \ctrl{3}\gategroup[wires=1,steps=1,style={draw=none,rounded corners, inner sep=1pt}, label style={label position=above}]{\footnotesize $G_{n}$}    & \qw\\
			\lstick{$\vdots$}  & \wave&&& \rstick{$\vdots$}\\
			\lstick{$q_{i_n}$}  & \qw &\ \ldots\ \qw & \qw  & \qw \\
			\lstick{$q_{j_1}$}  & \qw\gategroup[3,steps=3,style={draw=none,rounded corners},background,label style={label position=left,anchor=north,xshift=-1.4cm}]{\textit{P\textsubscript{J}}} &\ \ldots\ \qw  & \targ{}  & \qw \\
			\lstick{$\vdots$}  & \wave&&& \rstick{$\vdots$}\\
			\lstick{$q_{j_n}$}  & \targ{} &\ \ldots\ \qw & \qw  & \qw 
		\end{quantikz}
	\end{adjustbox}
 \label{fig5a}
	}
 \quad 
 \quad 
	\subfigure[]{  
		\begin{adjustbox}{width=0.28\textwidth}
			\begin{tikz}
				\draw [dashed,thick,red] (-3,1) -- (1.7,1);
			\end{tikz}
			\hspace{-5.2cm}
			\begin{quantikz}[row sep={0.6cm,between origins}]
				\lstick{$q_{i_1}$}  & \ctrl{5}\gategroup[3,steps=3,style={draw=none,rounded corners},background,label style={label position=left,anchor=north,xshift=-1.4cm}]{\textit{P\textsubscript{I}}}\gategroup[wires=1,steps=1,style={draw=none,rounded corners, inner sep=1pt}, label style={label position=above}]{\footnotesize $G_1$} &\ \ldots\ \qw & \ctrl{2}\gategroup[wires=1,steps=1,style={draw=none,rounded corners, inner sep=1pt}, label style={label position=above}]{\footnotesize $G_{n}$}    & \qw\\
				\lstick{$\vdots$}  & \wave&&& \rstick{$\vdots$}\\
				\lstick{$q_{i_n}$}  & \qw &\ \ldots\ \qw & \targ{}  & \qw \\
				\lstick{$q_{j_1}$}  & \qw\gategroup[3,steps=3,style={draw=none,rounded corners},background,label style={label position=left,anchor=north,xshift=-1.4cm}]{\textit{P\textsubscript{J}}} &\ \ldots\ \qw  & \qw  & \qw \\
				\lstick{$\vdots$}  & \wave&&& \rstick{$\vdots$}\\
				\lstick{$q_{j_n}$}  & \targ{} &\ \ldots\ \qw & \qw  & \qw 
			\end{quantikz}
		\end{adjustbox}
  \label{fig5b}
	} 
 \quad
 \quad 
	\subfigure[]{
		\begin{adjustbox}{width=0.28\textwidth}
			\begin{tikz}
				\draw [dashed,thick,red] (-3,1) -- (1.7,1);
			\end{tikz}
			\hspace{-5.2cm}
			\begin{quantikz}[row sep={0.6cm,between origins}]
				\lstick{$q_{i_1}$}  & \ctrl{5}\gategroup[3,steps=3,style={draw=none,rounded corners},background,label style={label position=left,anchor=north,xshift=-1.4cm}]{\textit{P\textsubscript{I}}}\gategroup[wires=1,steps=1,style={draw=none,rounded corners, inner sep=1pt}, label style={label position=above}]{\footnotesize $G_1$} &\ \ldots\ \qw & \qw\gategroup[wires=1,steps=1,style={draw=none,rounded corners, inner sep=1pt}, label style={label position=above}]{\footnotesize $G_{n}$}    & \qw\\
				\lstick{$\vdots$}  & \wave&&& \rstick{$\vdots$}\\
				\lstick{$q_{i_n}$}  & \qw &\ \ldots\ \qw & \ctrl{1}  & \qw \\
				\lstick{$q_{j_1}$}  & \qw\gategroup[3,steps=3,style={draw=none,rounded corners},background,label style={label position=left,anchor=north,xshift=-1.4cm}]{\textit{P\textsubscript{J}}} &\ \ldots\ \qw  & \targ{}  & \qw \\
				\lstick{$\vdots$}  & \wave&&& \rstick{$\vdots$}\\
				\lstick{$q_{j_n}$}  & \targ{} &\ \ldots\ \qw & \qw  & \qw 
			\end{quantikz}
		\end{adjustbox}
  \label{fig5c}
	}     
\caption{The different impacts of transferring qubit \(q_{i_1}\) on subsequent gates. (a), (b), and (c) indicate that the transfer of qubit  \(q_{i_1}\) have a positive impact, a negative impact, and no impact on \(G_n\), respectively.}
\label{fig5}

\end{figure*}

\begin{itemize}
\item The positive impact occurs when subsequent quantum gates are global gates belonging to the same partition as the current global gate, and one of the quantum states acts on the transmission qubit \(q_{i_1}\), as shown in Fig. \ref{fig5}(a). In this case, \(E_q\) is set to 1. The global gates are merged transferred to the target partition as a result of the current transmission qubit.
\item The negative impact occurs when subsequent quantum gates are local gates acting on the transmission qubit \(q_{i_1}\), as shown in Fig. \ref{fig5}(b). In this case, \(E_q\) is set to -1. The local gates can be executed within their own partition without the need for transmission, while unnecessary transmission increases in execution cost.
\item The no impact occurs when subsequent gates are global gates whose partition does not completely match the partition of the current global gate, or when subsequent gates do not act on the transmission qubit at all, as shown in Fig. \ref{fig5}(c). In this case, \(E_q\) is set to 0.
\end{itemize}

Therefore, before transferring the qubits for global gates, it is imperative to evaluate the impact of the selected transmission qubit on the transmission cost of subsequent gates. Ref. \cite{nikahd2021automated} has successfully reduced the transmission cost of distributed quantum circuits by introducing lookahead window technology. However, the lookahead window size set for each global gate transmission in these approaches is fixed. A too-large lookahead window might benefit later-executed quantum gates at the expense of earlier ones, thus increasing the overall cost. Conversely, a too-small lookahead window might fail to fully consider the impact on the transmission cost of subsequent gates. 

To overcome these limitations, a dynamic strategy is adopted to adjust the lookahead window size. When considering the transmission qubit for global gates, the lookahead window is flexibly set based on the relationship between subsequent quantum gates and the transmission qubit. To identify the optimal size for the lookahead window, a dynamic lookahead algorithm is proposed in Algorithm \ref{algorithm1}. 

\begin{algorithm}[h]
	\KwIn{the quantum circuit $QC$, the target gate $g$, the transmission qubit $q$}
	\KwOut{the lookahead window size $D_q$}
	\Begin{
		$g\_list  \leftarrow QC$\;
		$i \leftarrow $ \textnormal{index gate in $g\_list$}\;
		$gg\_list \leftarrow$ \textnormal{count global gate from $QC$}\; 
		$j \leftarrow i+1$\;
		\While{$j$ \textless $g\_list$ $length$}{
			\If{$g\_list[j]$ \textnormal{is in} $gg\_list$ and $g\_list[j]$ \textnormal{ is the same partition as} $g$ \textnormal{and $q$ is in $g\_list[i]$} }{
				$j \leftarrow i+1$; \tcp{positive impact}
			}
			\If{$g\_list[j]$ \textnormal{is not in} $gg\_list$ \textnormal{and $q$ is in $g\_list[i]$}}{
				break; \tcp{negative impact} 
			}
			\Else{ $j\leftarrow i+1$; \tcp{no impact}}
		}
		$D_q \leftarrow j-i+1$\;
		Return $D_q$\;
	}		
	
	\caption{Calculate the lookahead window size: CLAD ($QC$, $g$, $q$).}
	\label{algorithm1} 
\end{algorithm}

This algorithm takes a quantum circuit, the current global gate \(G_i\), and the transmission qubit \(q\) for the global gate as inputs and outputs the computed lookahead window size for the transmission qubit of the current global gate. For the current global gate \(G_i\) and its transmission qubit \(q\), the relationship between subsequent quantum gates and the transmission qubit \(q\) is evaluated in sequence until a gate \(G_j\) with a negative impact is identified. At this point, the index distance from the negatively impacting gate \(G_j\) to the current global gate \(G_i\), given by \(j-i+1\), is considered the lookahead window size \(D_q\) for the transmission qubit \(q\) of the current global gate \(G_i\).

Based on the optimal lookahead window size for the transmission qubit of the current global gate, it is necessary to assess the impact of different transmission qubits on the transmission cost of quantum gates in the subsequent circuit within the given lookahead window. This assessment aids in selecting the suitable transmission qubit. For this purpose, this paper evaluates the selection of the transmission qubit using the impact cost function \(F_q\) for the transmission qubit \(q\). The constructed function is presented in Eq. \eqref{9},
\begin{equation}
	F_q=\sum_{k=1}^{D_q-1} f_k=\sum_{k=1}^{D_q-1} (D_q-k)E_q, E_q \in \left \{ 0 ,\pm 1\right \} 
	\label{9}
\end{equation}
where \(f_k\) represents the impact cost function for each quantum gate within the subsequent lookahead window; \(D_q\) represents the lookahead window size for the transmission qubit \(q\); \(k\) represents the index of the quantum gates within the lookahead window, ranging from 0 to \(D_q-1\). The current global gate with index 0 does not need to calculate \(F_q\), so \(k\) starts from 1 in Eq. \eqref{9}. And \(E_q\) is the impact factor, indicating the impact of the current transmission qubit on the quantum gate indexed by \(k\), as previously described.

The value of the transmission qubit impact cost function \(F_q\) indicates the optimization effect of selecting the transmission qubit \(q\) of the current global gate on the transmission of subsequent quantum gates. A larger value signifies a better optimization effect based on transferring that particular qubit. This equation calculates the impact cost function \(f_k\) for each gate based on the relationship between the current transmission qubit and subsequent quantum gates. The smaller the value of \(k\), the closer it is to the current global gate. Due to the order of execution, quantum gates that are closer to the current global gate have a more significant impact on the cost function \(F_q\). Therefore, \(D_q-k\) is used as the weight for the impact factor.

\begin{figure*}[]
	\centering
	\setcounter{subfigure}{0}
	\subfigure[]{  
		\begin{adjustbox}{width=0.47\textwidth}		
			\begin{quantikz}[row sep={0.6cm,between origins},column sep=0.8cm]
				\lstick{$q_0$}  & \ctrl{3}\gategroup[3,steps=1,style={draw=none,rounded corners},background,label style={label position=left,anchor=north,xshift=-1.8cm}]{\textit{P\textsubscript{1}}} \gategroup[4,steps=1,style={dashed,rounded corners,fill=blue!40, inner xsep=2pt, inner ysep=-0.08cm, fill opacity=0.5},background,label style={label position=below,anchor=north,xshift=0.55cm,yshift=0.1cm}]{{\small $G_1$}} & \ctrl{4}\gategroup[5,steps=1,style={dashed,rounded corners,fill=green!40, inner xsep=2pt, inner ysep=-0.08cm, fill opacity=0.5},background,label style={label position=below,anchor=north,xshift=0.55cm,yshift=0.1cm}]{{\small $G_3$}} & \qw & \qw & \qw &\targ{} \gategroup[2,steps=1,style={dashed,rounded corners,fill=red!20, inner xsep=2pt, inner ysep=-0.08cm},background,label style={label position=below,anchor=north,xshift=0.55cm,yshift=0.1cm}]{{\small $G_8$}} &\ctrl{7} & \qw\\
				\lstick{$q_1$}  & \qw & \qw & \qw & \targ{}\gategroup[3,steps=1,style={dashed,rounded corners, inner xsep=2pt, inner ysep=-0.08cm},background,label style={label position=below,anchor=north,xshift=0.55cm,yshift=0.1cm}]{{\small $G_6$}} & \ctrl{3}\gategroup[4,steps=1,style={dashed,rounded corners, inner xsep=2pt, inner ysep=-0.08cm},background,label style={label position=below,anchor=north,xshift=0.55cm,yshift=0.1cm}]{{\small $G_7$}} &\ctrl{-1} & \qw & \qw    \\
				\lstick{$q_2$}  & \qw & \qw & \targ{}\gategroup[2,steps=1,style={dashed,rounded corners, inner xsep=2pt, inner ysep=-0.08cm},background,label style={label position=below,anchor=north,xshift=0.55cm,yshift=0.1cm}]{{\small $G_5$}} & \qw & \qw  &\qw & \qw & \qw  \\
				\lstick{$q_3$}  & \targ{}\gategroup[2,steps=1,style={draw=none,rounded corners},background,label style={label position=left,anchor=north,xshift=-1.8cm}]{\textit{P\textsubscript{2}}} & \qw & \ctrl{-1} & \ctrl{-2} & \qw &\ctrl{1} & \qw  & \qw \\
				\lstick{$q_4$}  & \qw & \targ{} & \qw & \qw & \targ{} &\targ{}\gategroup[wires=1,steps=1,style={draw=none,rounded corners, inner sep=1pt,inner ysep=-0.1cm}, label style={label position=below, label distance=0.3cm,xshift=0.55cm,yshift=0.23cm}]{\small $G_9$} & \qw  & \qw \\
				\lstick{$q_5$}  & \ctrl{2}\gategroup[3,steps=1,style={draw=none,rounded corners},background,label style={label position=left,anchor=north,xshift=-1.8cm}]{\textit{P\textsubscript{3}}}\gategroup[3,steps=1,style={dashed,rounded corners, inner xsep=2pt, inner ysep=-0.08cm},background,label style={label position=below,anchor=north,xshift=0.55cm,yshift=0.1cm}]{{\small $G_2$}} & \ctrl{1}\gategroup[2,steps=1,style={dashed,rounded corners, inner xsep=2pt, inner ysep=-0.08cm},background,label style={label position=below,anchor=north,xshift=0.55cm,yshift=0.1cm}]{{\small $G_4$}} & \qw & \qw & \qw &\qw & \qw  & \qw \\
				\lstick{$q_6$}  & \qw & \targ{} & \qw & \qw & \qw & \qw  &\qw & \qw \\
				\lstick{$q_7$}  & \targ{} & \qw & \qw & \qw & \qw &\qw & \targ{}\gategroup[wires=1,steps=1,style={draw=none,rounded corners, inner sep=1pt,inner ysep=-0.1cm}, label style={label position=below, label distance=0.3cm,xshift=0.55cm,yshift=0.23cm}]{\small $G_{10}$}  & \qw 
			\end{quantikz}
			\hspace{-10.39cm}
			\begin{tikzpicture}[>=stealth,baseline=0cm]
				\draw [dashed,thick,red] (-0.3,0.6) -- (9.9,0.6);
				\draw [dashed,thick,red] (-0.3,-0.6) -- (9.9,-0.6);
				\draw[fill=myorange!180] (2.12,2.1) circle (0.08cm);
				\draw[fill=purple!80] (2.12,2.1) circle (0.08cm);
				\node at (2.1,-0.5) [font=\normalsize] {$f = 0$};
				\node at (3.3,2.5) [font=\normalsize] {$f = 6$};
				\node at (3.27,-1.9) [font=\normalsize] {$f = 0$};
				\node at (4.4,1.3) [font=\normalsize] {$f = 0$};
				\node at (5.5,1.9) [font=\normalsize] {$f = 0$};
				\node at (6.6,1.9) [font=\normalsize] {$f = 0$};
				\node at (7.8,2.5) [font=\normalsize] {$f = -1$};	
                \coordinate (A) at (1.5,-2.45);
			\coordinate (B) at (1.5,2.7);
			\coordinate (C) at (8.5,2.7);
			\coordinate (D) at (8.5,0.8);
			\coordinate (E) at (7.4,0.8);
			\coordinate (F) at (7.4,-2.45);
                \node at (4.9,2.9) [font=\small] {Lookahead window size = 8};
			\draw[dashed,draw=gray!120, line width=2pt](A) -- (B) -- (C) -- (D) -- (E)-- (F)-- (A);
                \node at (4.9,-2.6) [font=\small] {};
			\end{tikzpicture}
   \label{22a}
		\end{adjustbox}
	}
\quad
	\subfigure[]{  
		\begin{adjustbox}{width=0.47\textwidth}

			\begin{quantikz}[row sep={0.6cm,between origins},column sep=0.8cm]
				\lstick{$q_0$}  & \ctrl{3}\gategroup[3,steps=1,style={draw=none,rounded corners},background,label style={label position=left,anchor=north,xshift=-1.8cm}]{\textit{P\textsubscript{1}}} \gategroup[4,steps=1,style={dashed,rounded corners,fill=blue!40, inner xsep=2pt, inner ysep=-0.08cm, fill opacity=0.5},background,label style={label position=below,anchor=north,xshift=0.55cm,yshift=0.1cm}]{{\small $G_1$}} & \ctrl{4}\gategroup[5,steps=1,style={dashed,rounded corners, inner xsep=2pt, inner ysep=-0.08cm},background,label style={label position=below,anchor=north,xshift=0.55cm,yshift=0.1cm}]{{\small $G_3$}} & \qw & \qw & \qw &\targ{} \gategroup[2,steps=1,style={dashed,rounded corners, inner xsep=2pt, inner ysep=-0.08cm},background,label style={label position=below,anchor=north,xshift=0.55cm,yshift=0.1cm}]{{\small $G_8$}} &\ctrl{7} & \qw\\
				\lstick{$q_1$}  & \qw & \qw & \qw & \targ{}\gategroup[3,steps=1,style={dashed,rounded corners,fill=green!40, inner xsep=2pt, inner ysep=-0.08cm, fill opacity=0.5},background,label style={label position=below,anchor=north,xshift=0.55cm,yshift=0.1cm}]{{\small $G_6$}} & \ctrl{3}\gategroup[4,steps=1,style={dashed,rounded corners, inner xsep=2pt, inner ysep=-0.08cm},background,label style={label position=below,anchor=north,xshift=0.55cm,yshift=0.1cm}]{{\small $G_7$}} &\ctrl{-1} & \qw & \qw    \\
				\lstick{$q_2$}  & \qw & \qw & \targ{}\gategroup[2,steps=1,style={dashed,rounded corners,fill=green!40, inner xsep=2pt, inner ysep=-0.08cm, fill opacity=0.5},background,label style={label position=below,anchor=north,xshift=0.55cm,yshift=0.1cm}]{{\small $G_5$}} & \qw & \qw  &\qw & \qw & \qw  \\
				\lstick{$q_3$}  & \targ{}\gategroup[2,steps=1,style={draw=none,rounded corners},background,label style={label position=left,anchor=north,xshift=-1.8cm}]{\textit{P\textsubscript{2}}} & \qw & \ctrl{-1} & \ctrl{-2} & \qw &\ctrl{1}\gategroup[2,steps=1,style={dashed,rounded corners,fill=red!40, inner xsep=2pt, inner ysep=-0.08cm, fill opacity=0.5},background,label style={label position=below,anchor=north,xshift=0.55cm,yshift=0.1cm}]{{\small $G_9$}} & \qw  & \qw \\
				\lstick{$q_4$}  & \qw & \targ{} & \qw & \qw & \targ{} &\targ{} & \qw  & \qw \\
				\lstick{$q_5$}  & \ctrl{2}\gategroup[3,steps=1,style={draw=none,rounded corners},background,label style={label position=left,anchor=north,xshift=-1.8cm}]{\textit{P\textsubscript{3}}}\gategroup[3,steps=1,style={dashed,rounded corners, inner xsep=2pt, inner ysep=-0.08cm},background,label style={label position=below,anchor=north,xshift=0.55cm,yshift=0.1cm}]{{\small $G_2$}} & \ctrl{1}\gategroup[2,steps=1,style={dashed,rounded corners, inner xsep=2pt, inner ysep=-0.08cm},background,label style={label position=below,anchor=north,xshift=0.55cm,yshift=0.1cm}]{{\small $G_4$}} & \qw & \qw & \qw &\qw & \qw  & \qw \\
				\lstick{$q_6$}  & \qw & \targ{} & \qw & \qw & \qw & \qw  &\qw & \qw \\
				\lstick{$q_7$}  & \targ{} & \qw & \qw & \qw & \qw &\qw & \targ{}\gategroup[wires=1,steps=1,style={draw=none,rounded corners, inner sep=1pt,inner ysep=-0.1cm}, label style={label position=below, label distance=0.3cm,xshift=0.55cm,yshift=0.23cm}]{\small $G_{10}$}  & \qw 
			\end{quantikz}
		\hspace{-10.376cm}
		\begin{tikzpicture}[>=stealth,baseline=0cm]
				\draw [dashed,thick,red] (-0.3,0.6) -- (9.9,0.6);
			\draw [dashed,thick,red] (-0.3,-0.6) -- (9.9,-0.6);
		\coordinate (A) at (1.5,-2.45);
		\coordinate (B) at (1.5,2.7);
		\coordinate (C) at (8.5,2.7);
		\coordinate (D) at (8.5,-2.45);
		\node at (4.9,2.9) [font=\small] {Lookahead window size = 9};
		\draw[dashed,draw=gray!120,line width=2pt](A) -- (B) -- (C) -- (D) -- (A);			
		\draw[fill=yellow!180] (2.1,0.294) circle (0.14cm);
		\draw[fill=red!180] (2.1,0.294) circle (0.14cm);
		\node at (2.1,-0.5) [font=\normalsize] {$f = 0$};
		\node at (3.3,2.5) [font=\normalsize] {$f = 0$};
		\node at (3.27,-1.9) [font=\normalsize] {$f = 0$};
		\node at (4.4,1.3) [font=\normalsize] {$f = 5$};
		\node at (5.5,1.9) [font=\normalsize] {$f = 4$};
		\node at (6.6,1.9) [font=\normalsize] {$f = 0$};
		\node at (7.8,2.5) [font=\normalsize] {$f = 0$};
		\node at (7.8,0.7) [font=\normalsize] {$f = -1$};
		\node at (4.9,-2.6) [font=\small] {};
	\end{tikzpicture}

   \label{22b}
			\end{adjustbox}
	} 
\caption{The impact of selecting different transmission qubits on the lookahead window size and the transmission cost. The green box, the red box, and the colorless box represent the transmission qubit having a positive impact, a negative impact, and no impact on the gate, respectively. (a) shows that selecting qubit \(q_0\) results in the lookahead window size \(D_{q_0} = 8\) and the impact cost function \(F_{q_0} = 5\); while (b) shows that selecting qubit \(q_3\) leads to \(D_{q_3} = 9\) and \(F_{q_3} = 8\).}
\label{222}
\end{figure*}
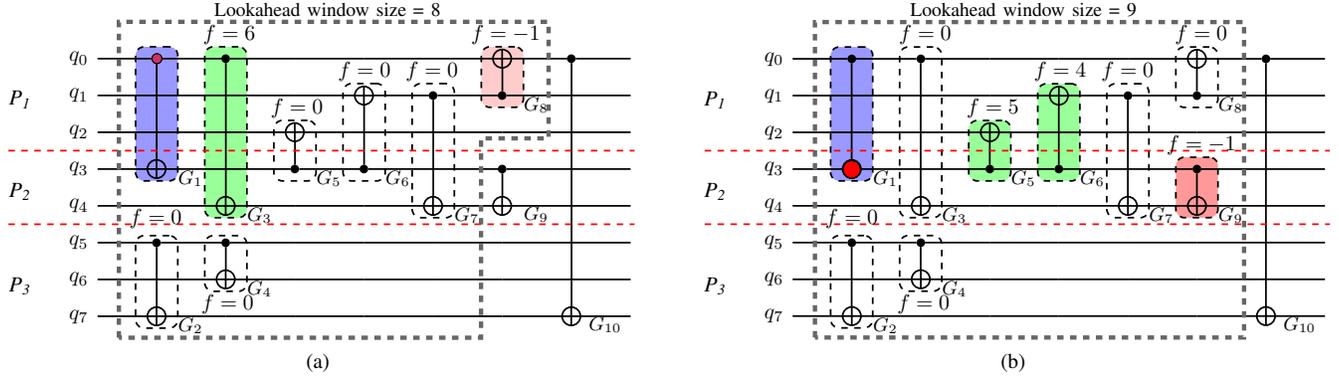

As shown in Fig. \ref{222}, there are different lookahead window sizes and impact cost functions for the two qubits of the gate \(G_1\). When the transmission qubit \(q_0\) is selected as shown in Fig. \ref{222}(a), it has no impact on \(G_2\), \(G_4-G_7\), a positive impact on \(G_3\), and a negative impact on \(G_8\). With \(G_0\) to \(G_7\) in the lookahead window \(D_{q_{0}}\), the impact cost function \(F_{q_0}\) for transferring qubit \(q_0\) is calculated as \((8-2)-(8-7)=5\). As shown in Fig. \ref{222}(b), when the transmission qubit \(q_3\) is selected, it has no impact on \(G_2-G_4\), \(G_7\), \(G_8\), a positive impact on \(G_5\), \(G_6\), and a negative impact on \(G_9\). With \(G_0\) to  \(G_8\) in the lookahead window \(D_{q_{3}}\), the impact cost function \(F_{q_3}\) for transferring qubit \(q_3\) is calculated as \((9-4)+(9-5)-(9-8)=8\). Therefore, the qubit \(q_3\) with the greater impact cost is selected as the transmission qubit.

\subsection{Transmission Cost Optimization Algorithm}
By dynamically selecting the lookahead window size for the transmission qubit of the current global gate, the transmission cost of distributed quantum circuits is optimized. To better describe the execution order of global gates, two variables are set during the optimization process: the merged transmission queue \(T\_{queue}\) and the merged transmission list \(T\_{list}\). The \(T\_{queue}\) is used for temporarily storing global gates that can be converted into local gates through a single transmission, i.e., a set of global gates that satisfy the merged transfer model. The \(T\_{list}\) is used to store the \(T\_{queue}\), with the $2|T\_{list}|$
 representing the transmission cost.

The main idea of the transmission cost optimization algorithm based on dynamic lookahead is as follows: First, a merged transmission queue \(T\_{queue}\) and a transmission list \(T\_{list}\) are initialized. Next, the lookahead window sizes \(D_c\) and \(D_t\) for the control and target qubits of each global gate are calculated, along with their respective impact cost functions \(F_c\) and \(F_t\). Then, based on these impact cost functions, the optimal transmission qubit is selected to match merged transfer model with the quantum gates within the lookahead window. Finally, global gates meeting the merged transfer condition are added to \(T\_{queue}\), which is then appended to \(T\_{list}\). This process is repeated until all global gates have been processed. The pseudocode is described in Algorithm \ref{algorithm2}.

\begin{algorithm}[h]
	\KwIn{the quantum circuit $QC$}
	\KwOut{the gate transmission list $T\_list$}
	\Begin{
		$g\_list  \leftarrow QC$\;
		$T\_list  \leftarrow [$ $]$\;
		\ForEach{$gate$ \textnormal{in} $g\_list$}{
			\If{$gate$ \textnormal{is global gate}}{
				$D_c \leftarrow $ CLAD ($QC$, $gate$, $control\  qubit$)\; 
                    $D_t \leftarrow $ CLAD ($QC$, $gate$, $target\ qubit$)\;
				$F_c,F_t \leftarrow $ Calcuate $F$ by $D_c,D_t$\;
                \If{$F_c$ \textgreater  $F_t$    }{
                $transmission\_qubit \leftarrow $ $F_c$\;}
                \Else{$transmission\_qubit \leftarrow $ $F_t$\;}
    		$T\_queue \leftarrow $ match merged transfer by $transmission\_qubit$ in $QC$\;
			}
			\Else{
				continue\;
			}
		}
		Add $T\_queue$ to $T\_list$\;
		Return $T\_list$\;
	}		
	
	\caption{Transmission cost optimization by dynamic lookahead (LA).}
	\label{algorithm2} 
\end{algorithm}
This algorithm leverages dynamic lookahead technology, and selects an appropriate lookahead window based on subsequent quantum gate conditions. It calculates the transmission qubits that can achieve better optimization effects in the transmission of subsequent quantum gates. By hierarchically considering the transmission of global and local gates, the algorithm can effectively optimize the transmission cost of distributed quantum circuits.

\section{Experimental Results and Analysis}\label{sec:5}
The algorithms in this paper are implemented in Python on an Intel 14900k CPU with 64GB RAM. The experiments utilized quantum circuits from the reversible circuit benchmark set RevLib and the Quantum Fourier Transform (QFT) algorithm, considering quantum circuits of various scales and structures. The transmission cost measured by the number of quantum state transmissions between partitions, serves as the performance indicator for comparing distributed quantum circuit implementations. The solution of the QUBO model in Sec. \ref{sec:3} is implemented using D-Wave's Advantage\_system4.1 quantum annealing computer with 5627 qubits and Hybrid Solver hybird\_binary\_quadratic\_model\_version2.

The experiments in this section are divided into four parts. In the first part, randomly generated circuits ranging from 10 qubits to 10,000 qubits are used to compare the runtime required by the proposed circuit partitioning method based on the QUBO model with other graph partitioning methods. In the second part, circuits from the RevLib are selected for testing. The dynamic lookahead method proposed is compared with the metaheuristic methods and greedy strategies in \cite{chen2023routing} and \cite{ghodsollahee2021connectivity} under two partition scenarios in terms of transmission cost. In the third part, QFT circuits are selected for testing, and the transmission cost optimization effects under multiple partition scenarios are compared with those in \cite{houshmand2020evolutionary,chen2023routing}
and \cite{daei2021improving}. The fourth part investigates the impact of load balancing tolerance on transmission cost during quantum circuit partitioning. Some related work, such as \cite{houshmand2020evolutionary}
and \cite{daei2021improving}, utilized quantum teleportation as the quantum state transmission technology in distributed quantum circuits, and the number of teleportations was used as the transmission cost. In contrast, this work, based on a superconducting distributed architecture, adopts direct quantum state transmission as the quantum state transmission technology and uses the number of direct transmissions as the transmission cost. Although the quantum state transmission technologies differ, the counting method remains consistent, ensuring the comparability of the experimental data.

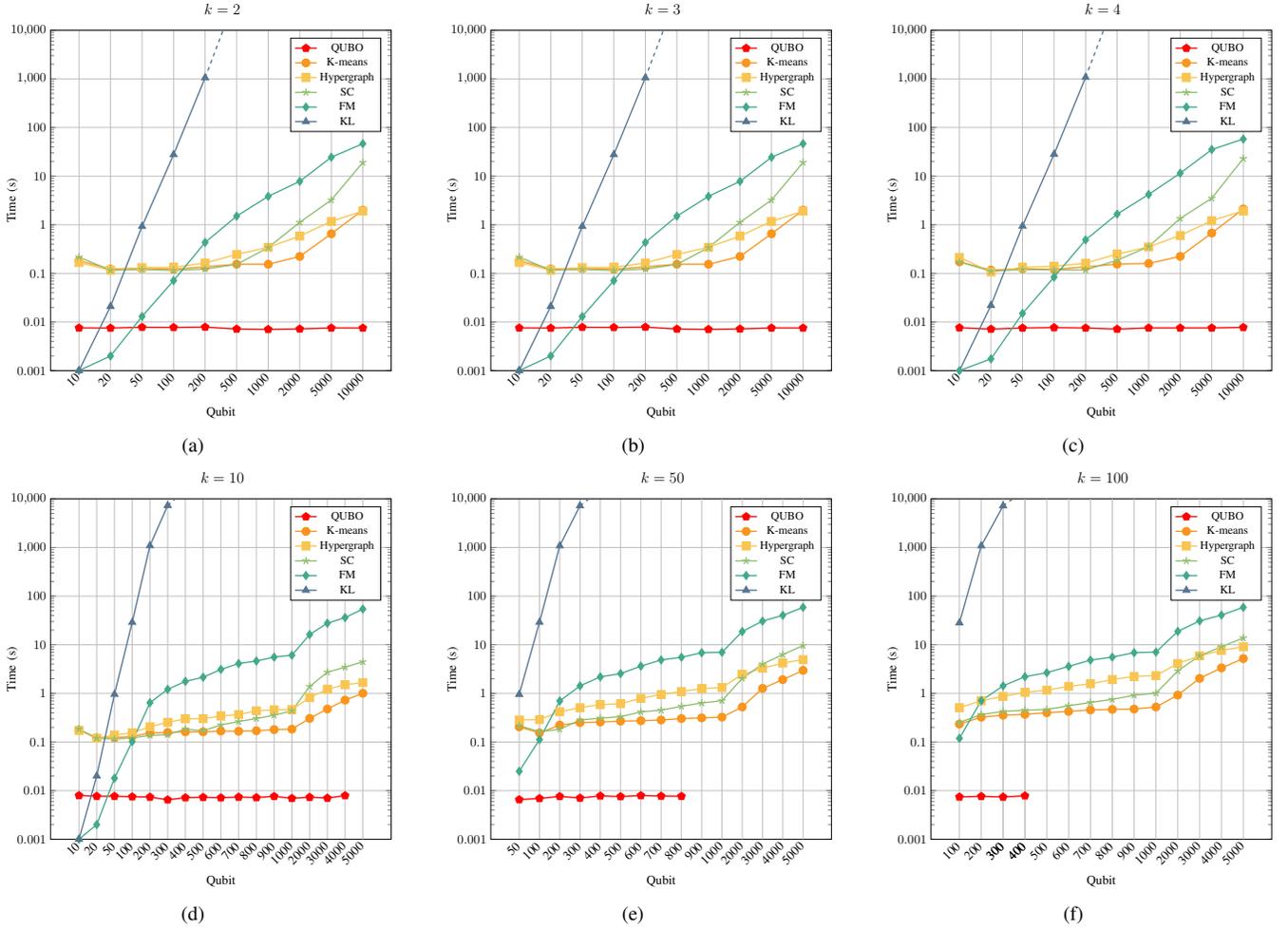
\begin{figure*}[h]
	\centering
	\setcounter{subfigure}{0}
	\subfigure[]{
        \begin{adjustbox}{width=0.3\textwidth}  
		\begin{tikzpicture}
  \node at(4.75,10) {\large$k = 2$};
			\begin{semilogyaxis}
				[xlabel={Qubit},
    ylabel={Time (s)},
    legend pos=north east,
    grid=major,
    symbolic x coords={10, 20, 50, 100, 200, 500, 1000, 2000, 5000, 10000}, 
    xtick=data, 
    log basis y={10},
    log ticks with fixed point,
    ymin=0.001, ymax=10000, 
    tick label style={/pgf/number format/fixed},
    width=11cm, height=11cm, 
    legend style={font=\small,yshift=2pt},
    xticklabel style={rotate=45, anchor=north east, yshift = 5pt},
    xlabel style={yshift=-10pt},
    ylabel style={yshift=-5pt},
    enlarge x limits=0.1, 
    xticklabel style={rotate=0}, 
    every x tick/.style={draw=none}, 
    cycle list name=color list, 
]
\addplot[color=pink, line width=1pt, mark=pentagon*, mark size=3pt] coordinates {
    (10, 0.006954) (20, 0.007882667) (50, 0.008026667) (100, 0.007781333)
    (200, 0.006897333) (500, 0.007446667) (1000, 0.006753333) (2000, 0.007241333)
    (5000, 0.006973333) (10000, 0.007576)
};
\addlegendentry{QUBO(Average)}
\addplot[color=red, line width=1pt,mark=square*,mark options={fill=red,rotate=45}, mark size=2pt] coordinates {
    (10, 0.020862) (20, 0.023648) (50, 0.02408) (100, 0.023344)
    (200, 0.020692) (500, 0.02234) (1000, 0.02026) (2000, 0.021724)
    (5000, 0.02092) (10000, 0.022728)
};
\addlegendentry{QUBO(Sum)}

\addplot[color=myorange, line width=1pt,mark=otimes*,mark options={fill=myorange}, mark size=3pt] coordinates {
    (10, 0.163002253) (20, 0.109003305) (50, 0.109001398) (100, 0.096997738)
    (200, 0.12800312) (500, 0.134004593) (1000, 0.151002169) (2000, 0.200003862)
    (5000, 0.566008091) (10000, 1.860912561)
};
\addlegendentry{K-means}

\addplot[color=mypurple,line width=1pt, mark=square*, mark size=3pt] coordinates {
    (10, 0.168003321) (20, 0.108002424) (50, 0.122002363) (100, 0.130002499)
    (200, 0.154002428) (500, 0.212002993) (1000, 0.313006401) (2000, 0.526007175)
    (5000, 1.009598017) (10000, 1.718923569)
};
\addlegendentry{Hypergraph}

\addplot[color=myred, line width=1pt,mark=star,mark options={fill=myred}, mark size=3pt] coordinates {
    (10, 0.155002594) (20, 0.091538191) (50, 0.103001356) (100, 0.113003731)
    (200, 0.112002134) (500, 0.139000177) (1000, 0.289002895) (2000, 1.105662584)
    (5000, 2.663036585) (10000, 19.9608016)
};
\addlegendentry{SC}

\addplot[color=mygreen,line width=1pt, mark=diamond*,mark options={fill=mygreen}, mark size=3pt] coordinates {
    (10, 0.00100112) (20, 0.002000332) (50, 0.010999918) (100, 0.056999207)
    (200, 0.225003242) (500, 0.766412497) (1000, 1.982553482) (2000, 4.035456419)
    (5000, 13.29440093) (10000, 31.26555467)
};
\addlegendentry{FM}

\addplot[color=myblue, line width=1pt,mark=triangle*,mark options={fill=myblue}, mark size=3pt] coordinates {
    (10, 0.001000643) (20, 0.019000053) (50, 0.912012339) (100, 26.403373) (200, 1050.541116)
};
\addlegendentry{KL}

\addplot[dashed,color=myblue, line width=1pt,mark=triangle*,mark options={fill=myblue},forget plot] coordinates {
    (200, 1050.541116) (500, 54000)
};

\end{semilogyaxis}

\end{tikzpicture}
\end{adjustbox}
	}
\quad
	\subfigure[]{
        \begin{adjustbox}{width=0.3\textwidth}  
		\begin{tikzpicture}
  \node at(4.75,10) {\large$k = 3$};
			\begin{semilogyaxis}
				[xlabel={Qubit},
    ylabel={Time (s)},
    legend pos=north east,
    grid=major,
    symbolic x coords={10, 20, 50, 100, 200, 500, 1000, 2000, 5000, 10000}, 
    xtick=data, 
    log basis y={10},
    log ticks with fixed point,
    ymin=0.001, ymax=10000, 
    tick label style={/pgf/number format/fixed},
    width=11cm, height=11cm, 
    legend style={font=\small,yshift=2pt},
    xticklabel style={rotate=45, anchor=north east, yshift = 5pt},
    xlabel style={yshift=-10pt},
    ylabel style={yshift=-5pt},
    enlarge x limits=0.1, 
    xticklabel style={rotate=0}, 
    every x tick/.style={draw=none}, 
    cycle list name=color list, 
]
\addplot[color=pink, line width=1pt, mark=pentagon*, mark size=3pt] coordinates {
    (10, 0.007154667) (20, 0.007698) (50, 0.00765) (100, 0.007546)
    (200, 0.007338) (500, 0.006738) (1000, 0.006086) (2000, 0.007594667)
    (5000, 0.007701333) (10000, 0.00736)
};
\addlegendentry{QUBO(Average)}
\addplot[color=red, line width=1pt,mark=square*,mark options={fill=red,rotate=45}, mark size=2pt] coordinates {
    (10, 0.021464) (20, 0.023094) (50, 0.02295) (100, 0.022638)
    (200, 0.022014) (500, 0.020214) (1000, 0.018258) (2000, 0.022784)
    (5000, 0.023104) (10000, 0.02208)
};
\addlegendentry{QUBO(Sum)}

\addplot[color=myorange, line width=1pt,mark=otimes*,mark options={fill=myorange}, mark size=3pt] coordinates {
    (10, 0.182999849) (20, 0.123000383) (50, 0.128001928) (100, 0.122002125)
    (200, 0.136002302) (500, 0.154543638) (1000, 0.154002905) (2000, 0.223001003)
    (5000, 0.653009176) (10000, 1.997611284)
};
\addlegendentry{K-means}

\addplot[color=mypurple,line width=1pt, mark=square*, mark size=3pt] coordinates {
    (10, 0.167217493) (20, 0.115999699) (50, 0.132000685) (100, 0.133003235)
    (200, 0.164002419) (500, 0.245561838) (1000, 0.344003201) (2000, 0.58600831)
    (5000, 1.174839973) (10000, 1.912106991)
};
\addlegendentry{Hypergraph}

\addplot[color=myred, line width=1pt,mark=star,mark options={fill=myred}, mark size=3pt] coordinates {
    (10, 0.217002153) (20, 0.116000891) (50, 0.120000839) (100, 0.115002394)
    (200, 0.123003006) (500, 0.153001785) (1000, 0.34006381) (2000, 1.122561455)
    (5000, 3.23060298) (10000, 19.10158014)
};
\addlegendentry{SC}

\addplot[color=mygreen,line width=1pt, mark=diamond*,mark options={fill=mygreen}, mark size=3pt] coordinates {
    (10, 0.001000643) (20, 0.0020012855529785156) (50, 0.01300025) (100, 0.070998192)
    (200, 0.433001041) (500, 1.506687641) (1000, 3.836079121) (2000, 7.811971903)
    (5000, 24.4376049) (10000, 46.45017958)
};
\addlegendentry{FM}

\addplot[color=myblue, line width=1pt,mark=triangle*,mark options={fill=myblue}, mark size=3pt] coordinates {
    (10, 0.001000166) (20, 0.021000385) (50, 0.928013325) (100, 27.72697473) (200, 1057.495415)
};
\addlegendentry{KL}

\addplot[dashed,color=myblue, line width=1pt,mark=triangle*,mark options={fill=myblue},forget plot] coordinates {
    (200, 1057.495415) (500, 54000)
};

\end{semilogyaxis}

\end{tikzpicture}
\end{adjustbox}
	}
 \quad
    \subfigure[]{
        \begin{adjustbox}{width=0.3\textwidth}  
		\begin{tikzpicture}
  \node at(4.75,10) {\large$k = 4$};
			\begin{semilogyaxis}
				[xlabel={Qubit},
    ylabel={Time (s)},
    legend pos=north east,
    grid=major,
    symbolic x coords={10, 20, 50, 100, 200, 500, 1000, 2000, 5000, 10000}, 
    xtick=data, 
    log basis y={10},
    log ticks with fixed point,
    ymin=0.001, ymax=10000, 
    tick label style={/pgf/number format/fixed},
    width=11cm, height=11cm, 
    legend style={font=\small,yshift=2pt},
    xticklabel style={rotate=45, anchor=north east, yshift = 5pt},
    xlabel style={yshift=-10pt},
    ylabel style={yshift=-5pt},
    enlarge x limits=0.1, 
    xticklabel style={rotate=0}, 
    every x tick/.style={draw=none}, 
    cycle list name=color list, 
]
\addplot[color=pink, line width=1pt, mark=pentagon*, mark size=3pt] coordinates {
    (10, 0.007706) (20, 0.007857333) (50, 0.007596667) (100, 0.007466)
    (200, 0.007204667) (500, 0.006396667) (1000, 0.006766) (2000, 0.007669333)
    (5000, 0.006991333) (10000, 0.007269333)
};
\addlegendentry{QUBO(Average)}
\addplot[color=red, line width=1pt,mark=square*,mark options={fill=red,rotate=45}, mark size=2pt] coordinates {
    (10, 0.023118) (20, 0.023572) (50, 0.02279) (100, 0.022398)
    (200, 0.021614) (500, 0.01919) (1000, 0.020298) (2000, 0.023008)
    (5000, 0.020974) (10000, 0.021808)
};
\addlegendentry{QUBO(Sum)}

\addplot[color=myorange, line width=1pt,mark=otimes*,mark options={fill=myorange}, mark size=3pt] coordinates {
    (10, 0.173003674) (20, 0.117001295) (50, 0.125239849) (100, 0.121006727)
    (200, 0.136002779) (500, 0.155002832) (1000, 0.160001993) (2000, 0.224001408)
    (5000, 0.675829411) (10000, 2.111235857)
};
\addlegendentry{K-means}

\addplot[color=mypurple,line width=1pt, mark=square*, mark size=3pt] coordinates {
    (10, 0.214003325) (20, 0.107004642) (50, 0.132999897) (100, 0.140002966)
    (200, 0.162004471) (500, 0.25000453) (1000, 0.350682974) (2000, 0.603008509)
    (5000, 1.218016386) (10000, 1.936426401)
};
\addlegendentry{Hypergraph}

\addplot[color=myred, line width=1pt,mark=star,mark options={fill=myred}, mark size=3pt] coordinates {
    (10, 0.172553062) (20, 0.111002922) (50, 0.120001793) (100, 0.118001938)
    (200, 0.1180019386) (500, 0.186001539) (1000, 0.356006622) (2000, 1.349581242)
    (5000, 3.511733532) (10000, 22.98588538)
};
\addlegendentry{SC}

\addplot[color=mygreen,line width=1pt, mark=diamond*,mark options={fill=mygreen}, mark size=3pt] coordinates {
    (10, 0.001000166) (20, 0.001746416) (50, 0.014998913) (100, 0.084001303)
    (200, 0.487949371) (500, 1.659022331) (1000, 4.206139565) (2000, 11.49473691)
    (5000, 35.15653968) (10000, 57.88318777)
};
\addlegendentry{FM}

\addplot[color=myblue, line width=1pt,mark=triangle*,mark options={fill=myblue}, mark size=3pt] coordinates {
    (10, 0.000999451) (20, 0.022000551) (50, 0.942648649) (100, 28.25496268) (200, 1078.962395)
};
\addlegendentry{KL}

\addplot[dashed,color=myblue, line width=1pt,mark=triangle*,mark options={fill=myblue},forget plot] coordinates {
    (200, 1078.962395) (500, 54000)
};

\end{semilogyaxis}
\end{tikzpicture}
\end{adjustbox}
	}
  \quad
    \subfigure[]{
        \begin{adjustbox}{width=0.3\textwidth}  
		\begin{tikzpicture}
  \node at(4.75,10) {\large$k = 10$};
			\begin{semilogyaxis}
				[xlabel={Qubit},
    ylabel={Time (s)},
    legend pos=north east,
    grid=major,
    symbolic x coords={10, 20, 50, 100, 200, 300, 400, 500, 600, 700, 800, 900, 1000, 2000, 3000, 4000, 5000}, 
    xtick=data, 
    extra x ticks={5000},
    log basis y={10},
    log ticks with fixed point,
    ymin=0.001, ymax=10000, 
    tick label style={/pgf/number format/fixed},
    width=11cm, height=11cm, 
    legend style={font=\small,yshift=2pt},
    xticklabel style={rotate=45, anchor=north east, yshift = 5pt},
    xlabel style={yshift=-10pt},
    ylabel style={yshift=-5pt},
    enlarge x limits=0.1, 
    xticklabel style={rotate=0}, 
    every x tick/.style={draw=none}, 
    cycle list name=color list, 
]
\addplot[color=pink, line width=1pt, mark=pentagon*, mark size=3pt] coordinates {
    (10, 0.007849333) (20, 0.007473333) (50, 0.007204) (100, 0.007449333)
    (200, 0.006412667) (300, 0.006086) (400, 0.006766) (500, 0.006894667)
    (600, 0.006393333) (700, 0.007952) (800, 0.007360667) (900, 0.007461333) (1000, 0.007402667) (2000, 0.007510667) (3000, 0.007769333) (4000, 0.007021333) 
};
\addlegendentry{QUBO(Average)}
\addplot[color=red, line width=1pt,mark=square*,mark options={fill=red,rotate=45}, mark size=2pt] coordinates {
    (10, 0.023548) (20, 0.02242) (50, 0.021612) (100, 0.022348)
    (200, 0.019238) (300, 0.018258) (400, 0.020298) (500, 0.020684)
    (600, 0.01918) (700, 0.023856) (800, 0.022082) (900, 0.022384) (1000, 0.022208) (2000, 0.022532) (3000, 0.023308) (4000, 0.021064) 
};
\addlegendentry{QUBO(Sum)}

\addplot[color=myorange, line width=1pt,mark=otimes*,mark options={fill=myorange}, mark size=3pt] coordinates {
    (10, 0.176002741) (20, 0.122001886) (50, 0.123002529) (100, 0.131002426)
    (200, 0.153002739) (300, 0.157550812) (400, 0.162001848) (500, 0.162003756)
    (600, 0.168003082) (700, 0.166554928) (800, 0.170555592) (900, 0.180003405) (1000, 0.184002638) (2000, 0.306004047) (3000, 0.478006124) (4000, 0.727010727) (5000, 1.005552769)
};
\addlegendentry{K-means}

\addplot[color=mypurple,line width=1pt, mark=square*, mark size=3pt] coordinates {
    (10, 0.173996687) (20, 0.120002508) (50, 0.139552593) (100, 0.155003309)
    (200, 0.206003428) (300, 0.254206657) (400, 0.300005674) (500, 0.302008152)
    (600, 0.342004299) (700, 0.36756134) (800, 0.439006567) (900, 0.460008621) (1000, 0.466008425) (2000, 0.820012331) (3000, 1.219481945) (4000, 1.506293774) (5000, 1.668734789)
};
\addlegendentry{Hypergraph}

\addplot[color=myred, line width=1pt,mark=star,mark options={fill=myred}, mark size=3pt] coordinates {
    (10, 0.186002016) (20, 0.119997263) (50, 0.115002871) (100, 0.122002125)
    (200, 0.137002707) (300, 0.142000198) (400, 0.184565306) (500, 0.170557261)
    (600, 0.224743605) (700, 0.264005184) (800, 0.305004358) (900, 0.355004787) (1000, 0.439191103) (2000, 1.385409832) (3000, 2.72120595) (4000, 3.4549453267) (5000, 4.447271585)
};
\addlegendentry{SC}

\addplot[color=mygreen,line width=1pt, mark=diamond*,mark options={fill=mygreen}, mark size=3pt] coordinates {
    (10, 0.001000643) (20, 0.001999855) (50, 0.018000126) (100, 0.102001429)
    (200, 0.64200902) (300, 1.210941315) (400, 1.766746044) (500, 2.144617319)
    (600, 3.108271599) (700, 4.10600996) (800, 4.632513046) (900, 5.568226337) (1000, 6.088901281) (2000, 16.19960666) (3000, 27.67763877) (4000, 36.25393271) (5000, 54.21181345)
};
\addlegendentry{FM}

\addplot[color=myblue, line width=1pt,mark=triangle*,mark options={fill=myblue}, mark size=3pt] coordinates {
    (10, 0.001000166) (20, 0.019999266) (50, 0.957596779) (100, 28.90125799) (200, 1090.637719)
    (300, 7293.61858)
};
\addlegendentry{KL}

\addplot[dashed,color=myblue, line width=1pt,mark=triangle*,mark options={fill=myblue},forget plot] coordinates {
    (300, 7293.61858) (400, 15000)
};

\end{semilogyaxis}
\end{tikzpicture}
\end{adjustbox}
	}
   \quad
    \subfigure[]{
        \begin{adjustbox}{width=0.3\textwidth}  
		\begin{tikzpicture}
  \node at(4.75,10) {\large$k = 50$};
			\begin{semilogyaxis}
				[xlabel={Qubit},
    ylabel={Time (s)},
    legend pos=north east,
    grid=major,
    symbolic x coords={50, 100, 200, 300, 400, 500, 600, 700, 800, 900, 1000, 2000, 3000, 4000, 5000}, 
    xtick=data, 
    extra x ticks={900, 1000, 2000, 3000, 4000, 5000},
    log basis y={10},
    log ticks with fixed point,
    ymin=0.001, ymax=10000, 
    tick label style={/pgf/number format/fixed},
    width=11cm, height=11cm, 
    legend style={font=\small,yshift=2pt},
    xticklabel style={rotate=45, anchor=north east, yshift = 5pt},
    xlabel style={yshift=-10pt},
    ylabel style={yshift=-5pt},
    enlarge x limits=0.1, 
    xticklabel style={rotate=0}, 
    every x tick/.style={draw=none}, 
    cycle list name=color list, 
]
\addplot[color=pink, line width=1pt, mark=pentagon*, mark size=3pt] coordinates {
    (50, 0.006788) (100, 0.007078667)
    (200, 0.007360667) (300, 0.007725333) (400, 0.007325333) (500, 0.0082)
    (600, 0.007981333) (700, 0.007523333) (800, 0.007740667) 
};
\addlegendentry{QUBO(Average)}
\addplot[color=red, line width=1pt,mark=square*,mark options={fill=red,rotate=45}, mark size=2pt] coordinates {
    (50, 0.020364) (100, 0.021236)
    (200, 0.022082) (300, 0.023176) (400, 0.021976) (500, 0.0246)
    (600, 0.023944) (700, 0.02257) (800, 0.023222) 
};
\addlegendentry{QUBO(Sum)}

\addplot[color=myorange, line width=1pt,mark=otimes*,mark options={fill=myorange}, mark size=3pt] coordinates {
    (50, 0.204999924) (100, 0.15400362)
    (200, 0.223539352) (300, 0.251003981) (400, 0.258013725) (500, 0.269003868)
    (600, 0.274478436) (700, 0.284004688) (800, 0.303551674) (900, 0.31400466) (1000, 0.325004339) (2000, 0.525007248) (3000, 1.268116474) (4000, 1.917026043) (5000, 2.971981525)
};
\addlegendentry{K-means}

\addplot[color=mypurple,line width=1pt, mark=square*, mark size=3pt] coordinates {
    (50, 0.284004211) (100, 0.289004326)
    (200, 0.420006275) (300, 0.509006977) (400, 0.587430716) (500, 0.615009546)
    (600, 0.79001236) (700, 0.95286417) (800, 1.092584133) (900, 1.25201869) (1000, 1.319359779) (2000, 2.444659233) (3000, 3.290442944) (4000, 4.207123041) (5000, 4.950179577)
};
\addlegendentry{Hypergraph}

\addplot[color=myred, line width=1pt,mark=star,mark options={fill=myred}, mark size=3pt] coordinates {
    (50, 0.217000246) (100, 0.163002491)
    (200, 0.183002949) (300, 0.28400445) (400, 0.310003757) (500, 0.332003832)
    (600, 0.417004824) (700, 0.452007532) (800, 0.542562246) (900, 0.635011911) (1000, 0.707010984) (2000, 2.039763451) (3000, 4.000165224) (4000, 6.274936914) (5000, 9.62346983)
};
\addlegendentry{SC}

\addplot[color=mygreen,line width=1pt, mark=diamond*,mark options={fill=mygreen}, mark size=3pt] coordinates {
    (50, 0.025002003) (100, 0.112001419)
    (200, 0.69922924) (300, 1.427605629) (400, 2.183236599) (500, 2.548620939)
    (600, 3.632348299) (700, 4.863576651) (800, 5.528870344) (900, 6.846395731) (1000, 6.992348433) (2000, 18.68330216) (3000, 30.64399338) (4000, 40.25088143) (5000, 58.79996705)
};
\addlegendentry{FM}

\addplot[color=myblue, line width=1pt,mark=triangle*,mark options={fill=myblue}, mark size=3pt] coordinates {
    (50, 0.952921867) (100, 28.99775958) (200, 1084.946105)
    (300, 7226.41636)
};
\addlegendentry{KL}

\addplot[dashed,color=myblue, line width=1pt,mark=triangle*,mark options={fill=myblue},forget plot] coordinates {
    (300, 7226.41636) (400, 15000)
};

\end{semilogyaxis}
\end{tikzpicture}
\end{adjustbox}
	}
 \quad
    \subfigure[]{
        \begin{adjustbox}{width=0.3\textwidth} 
		\begin{tikzpicture}
          \node at(4.75,10) {\large$k = 100$};
			\begin{semilogyaxis}
				[xlabel={Qubit},
    ylabel={Time (s)},
    legend pos=north east,
    grid=major,
    symbolic x coords={100, 200, 300, 400, 500, 600, 700, 800, 900, 1000, 2000, 3000, 4000, 5000}, 
    xtick=data, 
    extra x ticks={300, 400, 500, 600, 700, 800, 900, 1000, 2000, 3000, 4000, 5000},
    log basis y={10},
    log ticks with fixed point,
    ymin=0.001, ymax=10000, 
    tick label style={/pgf/number format/fixed},
    width=11cm, height=11cm, 
    legend style={font=\small,yshift=2pt},
    xticklabel style={rotate=45, anchor=north east, yshift = 5pt},
    xlabel style={yshift=-10pt},
    ylabel style={yshift=-5pt},
    enlarge x limits=0.1, 
    xticklabel style={rotate=0}, 
    every x tick/.style={draw=none}, 
    cycle list name=color list, 
]
\addplot[color=pink, line width=1pt, mark=pentagon*, mark size=3pt] coordinates {
    (100, 0.008570667)
    (200, 0.006792667) (300, 0.006893333) (400, 0.007154667) 
};
\addlegendentry{QUBO(Average)}
\addplot[color=red, line width=1pt,mark=square*,mark options={fill=red,rotate=45}, mark size=2pt] coordinates {
    (100, 0.025712)
    (200, 0.020378) (300, 0.02068) (400, 0.021464) 
};
\addlegendentry{QUBO(Sum)}

\addplot[color=myorange, line width=1pt,mark=otimes*,mark options={fill=myorange}, mark size=3pt] coordinates {
    (100, 0.234002829)
    (200, 0.324094534) (300, 0.357005119) (400, 0.371007681) (500, 0.400006294)
    (600, 0.426012993) (700, 0.457007885) (800, 0.46856308) (900, 0.476006985) (1000, 0.525007486) (2000, 0.922017813) (3000, 2.033579588) (4000, 3.344565392) (5000, 5.193157911)
};
\addlegendentry{K-means}

\addplot[color=mypurple,line width=1pt, mark=square*, mark size=3pt] coordinates {
    (100, 0.510541916)
    (200, 0.705010414) (300, 0.870383978) (400, 1.050014973) (500, 1.169017553)
    (600, 1.393630981) (700, 1.577023029) (800, 1.921597719) (900, 2.233801126) (1000, 2.314593792) (2000, 4.110662699) (3000, 5.913236618) (4000, 7.631101131) (5000, 9.05559516)
};
\addlegendentry{Hypergraph}

\addplot[color=myred, line width=1pt,mark=star,mark options={fill=myred}, mark size=3pt] coordinates {
    (100, 0.253004551)
    (200, 0.366217852) (300, 0.427006721) (400, 0.451007366) (500, 0.466558456)
    (600, 0.55700922) (700, 0.65100956) (800, 0.753558636) (900, 0.917014599) (1000, 1.009015322) (2000, 2.896630764) (3000, 5.93986845) (4000, 9.07002449) (5000, 13.80938029)
};
\addlegendentry{SC}

\addplot[color=mygreen,line width=1pt, mark=diamond*,mark options={fill=mygreen}, mark size=3pt] coordinates {
    (100, 0.119000435)
    (200, 0.707587242) (300, 1.430016041) (400, 2.201263666) (500, 2.660274982)
    (600, 3.585178375) (700, 4.830685616) (800, 5.564875841) (900, 6.876481533) (1000, 7.062817812) (2000, 18.81757808) (3000, 30.94526696) (4000, 40.77480459) (5000, 58.79298115)
};
\addlegendentry{FM}

\addplot[color=myblue, line width=1pt,mark=triangle*,mark options={fill=myblue}, mark size=3pt] coordinates {
    (100, 28.31488991) (200, 1083.574505)
    (300, 7192.319525)
};
\addlegendentry{KL}

\addplot[dashed,color=myblue, line width=1pt,mark=triangle*,mark options={fill=myblue},forget plot] coordinates {
    (300, 7192.319525) (400, 15000)
};

\end{semilogyaxis}
\end{tikzpicture}
\end{adjustbox}
	}
\caption{Comparison of runtime performance between the QUBO model-based and other circuit partitioning methods under different numbers of qubits and partition counts $k$. (a)-(f) respectively show the runtime and trends of various circuit partitioning methods for scenarios with $k$ ranging from 2 to 100.}

\label{fig1122}
\end{figure*}

To investigate the runtime advantages of the proposed circuit partitioning method based on the QUBO model, five commonly used graph partitioning algorithms are selected for comparison. These include K-means clustering, hypergraph partitioning, spectral clustering (SC), the Fiduccia-Mattheyses (FM) algorithm, and the Kernighan-Lin (KL) algorithm, all of which have been widely adopted in \cite{yan2023fuzzy,daei2020optimized,cambiucci2023hypergraphic,andres2019automated}. The comparison results are shown in Fig. \ref{fig1122}. Since Dwave's Hybrid Solver requires multiple calls to the quantum annealing computer for execution, the QUBO time contains the total and average execution time of multiple QPUs.

As depicted in Fig. \ref{fig1122}, the QUBO-based partitioning method demonstrates a clear advantage in runtime when handling quantum circuits of various sizes. Whether the number of qubits in the circuit increases from 10 to 10,000, the QUBO method maintains a stable runtime of approximately 0.007 seconds, achieving millisecond-level performance and is nearly unaffected by the problem size. This short runtime is attributed to the accelerated performance of quantum computing. In contrast, the runtimes of the other five partitioning methods increase with the number of qubits, significantly surpassing that of the QUBO method. This indicates that the circuit partitioning method based on the QUBO model offers strong stability and efficiency across different problem scales.

Additionally, the experiments examine the runtime performance of each method under different partition counts, $k$. As $k$ increases, the demand for physical qubits in the quantum annealer required by the QUBO method rises rapidly. The number of decision variables $kn$ (where $n$ represents the number of qubits in the circuit to be partitioned) in the QUBO method represents the required physical qubits. When $k$ reaches 50 or 100, this number exceeds the capacity of physical qubits and couplers in the annealer for larger quantum circuits. As a result, the QUBO method becomes infeasible to implement on the annealer in such cases, as shown in Figs. \ref{fig1122}(e) and \ref{fig1122}(f), where the QUBO method only yields results for a limited number of qubits. However, the runtime of the other partitioning methods grows significantly as the qubit count increases. Despite the hardware limitations faced by the QUBO-based partitioning method in high-partition and large-scale cases, it still holds the potential to outperform traditional graph partitioning methods in terms of runtime, especially for partitioning large quantum circuits in most practical scenarios.

To evaluate the optimization effect of the lookahead method (LA) on the transmission cost,  the LA method is compared with the results from \cite{chen2023routing} and \cite{ghodsollahee2021connectivity}. The experimental results are shown in Table \ref{table1} and \ref{table333}, where \textit{Circuit} represents the name of the quantum circuit. \textit{Qubit} represents the number of qubits required for the circuit. \textit{Gate} represents the number of two-qubit gates in the circuit. \textit{K} represents the number of partitions. \textit{TC} represents the transmission cost required to execute this distributed quantum circuit.  \textit{Time} in Table \ref{table1} represents the runtime in seconds to calculate the transmission cost, where \textit{Time[LA]} is the runtime of Algorithm \ref{algorithm2}. \textit{Imp} represents the optimization rate of the LA method in terms of transmission cost compared with the comparison results. \textit{Speedup} in Table \ref{table1} represents the improvement in runtime of the LA method compared with \cite{chen2023routing}.

\begin{table*}[]
\caption{Comparison of transmission cost and runtime with results from \cite{chen2023routing}.}
\centering
\footnotesize
\begin{tabular}{@{}ccccccccc@{}}
\toprule
\textbf{Circuit} & \textbf{Qubit} & \textbf{Gate} & \textbf{TC{\cite{chen2023routing}}} & \textbf{Time{\cite{chen2023routing}}} & \textbf{TC{[}LA{]}} & \textbf{Time{[}LA{]}} & \textbf{Imp} & \textbf{Speedup}\\ \midrule
4gt12-v0\_87     & 6              & 112           & 19                & 176.845             & 18                & 0.007              & 10.53\%      &25264\\
4gt4-v0\_72      & 6              & 113           & 20                & 193.767             & 20                & 0.009              & 0.00\%       &21530\\
alu-v2\_30        & 6              & 223           & 46                & 729.161             & 48                & 0.025              & -4.35\%      &25264\\
mod5adder\_127   & 6              & 239           & 47                & 900.035             & 47                & 0.029              & 0.00\%       &29166\\
sf\_274          & 6              & 336           & 44                & 1320.224            & 44                & 0.046              & 0.00\%       &31036\\
hwb5\_53         & 6              & 598           & 130                & 5164.638            & 132                & 0.219              & -1.43\%      &23853 \\
rd53\_138        & 8              & 60            & 7                 & 53.617              & 6                 & 0.003              & 14.29\%      &17872\\
cm82a\_208       & 8              & 283           & 23                & 1474.978            & 22                & 0.022              & 4.35\%       &67044\\
rd53\_251        & 8              & 564           & 101               & 5941.854            & 81                & 0.139              & 19.80\%      &42747\\
ising\_model\_10 & 10             & 90            & 10                & 101.022             & 10                & 0.003              & 0.00\%       &33674\\
mini\_alu\_305   & 10             & 77            & 7                & 100.546             & 7                & 0.003              & 0.00\%       &33515\\
rd73\_140        & 10             & 104           & 11                & 165.066             & 11                & 0.007              & 0.00\%       &23581\\
sys6-v0\_111     & 10             & 98            & 11                & 141.951             & 10                & 0.006              & 0.00\%       &23659\\
ham7\_299        & 21             & 151           & 31                & 944.498             & 14                & 0.008              & 54.84\%      &118062\\
sym9\_317        & 27             & 240           & 51                & 1696.989            & 14                & 0.019              & 72.55\%      &89315\\
cycle10\_293     & 39             & 222           & 65                & 3282.944            & 17                & 0.018              & 73.85\%      &182386\\
ham15\_298       & 45             & 313           & 64                & 6899.038            & 29                & 0.028              & 54.69\%      &246394\\ \bottomrule
\end{tabular}
\label{table1}
\end{table*}

\begin{table}[]
\caption{Comparison of transmission cost with the results from  \cite{ghodsollahee2021connectivity}.}
\centering
\begin{tabular}{@{}ccccccc@{}}
\toprule
\textbf{Circuit}        & \textbf{Qubit} & \textbf{Gate} & \textbf{K} & \textbf{TC{}\cite{ghodsollahee2021connectivity}{}} & \textbf{TC{[}LA{]}} & \textbf{Imp}     \\ \midrule
4gt5\_76       & 5     & 46   & 2 & 8          & 8          & 0.00\%  \\
4moudlo7       & 5     & 36   & 2 & 10         & 6          & 40.00\% \\
sym9\_147      & 6     & 114  & 2 & 16         & 10         & 37.50\% \\
mini\_alu\_305 & 9     & 77   & 2 & 18         & 8          & 55.56\% \\
sym6\_316      & 14    & 123  & 2 & 16         & 8          & 50.00\% \\
parity\_247    & 17    & 15   & 2 & 2          & 2          & 0.00\%  \\
ham7\_299      & 21    & 151  & 2 & 42         & 24         & 42.86\% \\ \bottomrule
\end{tabular}
\label{table333}
\end{table}

It should be noted that since the need to reduce the quantum state re-transmission back to the original partition is considered in \cite{chen2023routing}, the transmission cost is not $2|T\_{list}|$ in \textit{TC}\cite{chen2023routing}. Specifically, if no further quantum gate operations require quantum state transmission after the last global gate transmission is completed, the quantum state of the last global gate does not need to be re-transferred back to the original partition. To ensure comparability, \textit{TC[LA]} in Table \ref{table1} also takes this factor into account. 

From Table \ref{table1}, it is evident that for the majority of benchmark circuit tests, the LA method proposed outperforms the comparative method in terms of optimizing transmission cost and runtime, achieving an average optimization rate of 18.12\% and a peak rate of 73.85\% in transmission cost. The genetic algorithm used in \cite{chen2023routing} selects the transmission direction of global gates to optimize the transmission cost. However, due to limitations such as population size and the number of iterations, it is possible to find the optimal transmission direction of the global gate in a circuit with fewer qubits, but it is challenging in a circuit with more qubits. In contrast, the LA method proposed  intelligently selects the transmission qubits for global gates, avoiding local optima and achieving significant optimization results in circuits with more qubits, such as ham7\_299, sym9\_317, cycle10\_293 and ham15\_298. Moreover, in terms of computational efficiency, for circuits containing \(n\) global gates, even though the search space for transmission qubits is \(2^n\), the LA method only needs \(2n\) evaluations to reach a better solution. Compared with the genetic algorithm in \cite{chen2023routing} which takes at least tens of seconds, the LA method can be completed within 1 second, with an average speedup of 60866$\times$. This approach reduces computational complexity and enhances practical usability.

Table \ref{table333} presents a comparison of experimental results between the LA method and the method from \cite{ghodsollahee2021connectivity}, showing an average optimization rate of 32.27\% and a peak rate of 55.56\%. Ref. \cite{ghodsollahee2021connectivity} abstracts distributed quantum circuits into a matrix model, adopting a greedy strategy that merges adjacent matrices to reduce transmission cost. In contrast, the LA method utilizes lookahead technology to consider not just the gates adjacent to the current quantum gate but also the impact of subsequent gates, choosing gates with a more optimal cost to merge. By dynamically selecting the lookahead window, the LA method can more specifically optimize, reduce the search space, and improve search efficiency.

To verify the optimization effect of the LA method across multiple partitions, the QFT circuits are selected for experimental comparison with \cite{houshmand2020evolutionary,chen2023routing} and \cite{daei2021improving}, as shown in Table \ref{table2}, with an average optimization rate of 10.20\%. The transmission cost in  \cite{daei2021improving} and the LA method are significantly better than those methods described in \cite{houshmand2020evolutionary,chen2023routing}. Ref. \cite{daei2021improving} also employs lookahead technology and uses movement rules to advance subsequent gates that satisfy the merged transfer model within the circuit. For distributed circuits with more than two partitions, both \cite{daei2021improving} and this work assume a fully connected state between partitions, allowing for direct interactions. Compared with \cite{daei2021improving}, which looks ahead to the last global quantum gate in the circuit, the LA method only needs to look ahead to negatively impacted gates, resulting in a smaller lookahead window. This not only helps reduce computational complexity but also more effectively addresses the interrelationships between quantum gates. 

\begin{table}[]
\caption{Comparison of transmission cost for QFT circuits across multiple partitions with the results from \cite{houshmand2020evolutionary,chen2023routing} and \cite{daei2021improving}.}
\centering
\setlength{\tabcolsep}{4pt}
\begin{tabular}{@{}cccccccc@{}}
\toprule
\textbf{Circuit} & \textbf{Qubit} & \textbf{K} & \textbf{TC\cite{houshmand2020evolutionary}} & \textbf{TC\cite{chen2023routing}} & \textbf{TC\cite{daei2021improving}} & \textbf{TC[LA]} & \textbf{Imp} \\
\midrule
\multirow{3}{*}{4\_QFT}  & \multirow{3}{*}{4}     & 2 & 8       & 4       & 4       & 4       & 0.00\%  \\
                         &       & 3 & NA      & NA      & 10      & 6       & 40.00\% \\
                         &       & 4 & NA      & NA      & 14      & 12      & 14.29\% \\
\multirow{3}{*}{8\_QFT}  & \multirow{3}{*}{8}     & 2 & 38      & 12      & 8       & 8       & 0.00\%  \\
                         &       & 3 & NA      & NA      & 20      & 14      & 30.00\% \\
                         &       & 4 & NA      & NA      & 24      & 24      & 0.00\%  \\
\multirow{3}{*}{16\_QFT} & \multirow{3}{*}{16}    & 2 & 132     & 26      & 16      & 16      & 0.00\%  \\
                         &       & 3 & NA      & NA      & 40      & 30      & 25.00\% \\
                         &       & 4 & NA      & NA      & 48      & 48      & 0.00\%  \\
\multirow{3}{*}{32\_QFT} & \multirow{3}{*}{32}    & 2 & 532     & 65      & 32      & 32      & 0.00\%  \\
                         &       & 3 & NA      & NA      & 80      & 62      & 22.50\% \\
                         &       & 4 & NA      & NA      & 96      & 96      & 0.00\%  \\
\multirow{3}{*}{64\_QFT} & \multirow{3}{*}{64}    & 2 & 2250    & 155     & 64      & 64      & 0.00\%  \\
                         &       & 3 & NA      & NA      & 160     & 126     & 21.25\% \\
                         &       & 4 & NA      & NA      & 192     & 192     & 0.00\%  \\
\bottomrule
\end{tabular}
\label{table2}

\end{table}

\begin{figure}
\begin{adjustbox}{width=0.45\textwidth}  
	\begin{tikzpicture}
     	\begin{axis}[
			ybar,
			bar width=0.3cm,
                width=10cm, 
                height=6.18cm, 
			xlabel={},
			ylabel={TC (log scale)},
			legend style={at={(0.5,-0.15)},
				anchor=north,legend columns=-1},
			symbolic x coords={QFT4,QFT8,QFT16,QFT32,QFT64},
			xtick=data,
			ymode=log,
			log basis y=10,
			]  
			\addplot coordinates { (QFT4, 8) (QFT8, 38) (QFT16, 132) (QFT32, 532) (QFT64, 2250) };
			\addplot coordinates { (QFT4, 4) (QFT8, 12) (QFT16, 26) (QFT32, 65) (QFT64, 155) };
			\addplot coordinates { (QFT4, 4) (QFT8, 8) (QFT16, 16) (QFT32, 32) (QFT64, 64) };
			\addplot coordinates { (QFT4, 4) (QFT8, 8) (QFT16, 16) (QFT32, 32) (QFT64, 64) };
			\legend{Ref. \cite{houshmand2020evolutionary}, Ref. \cite{chen2023routing}, Ref. \cite{daei2021improving}, LA}
		\end{axis}
	\end{tikzpicture}
        \end{adjustbox}
	\caption{Bar chart comparing the transmission cost results in QFT circuits with  \cite{houshmand2020evolutionary,chen2023routing} and \cite{daei2021improving}.}
\label{fig:bar_chart_log}
\end{figure}
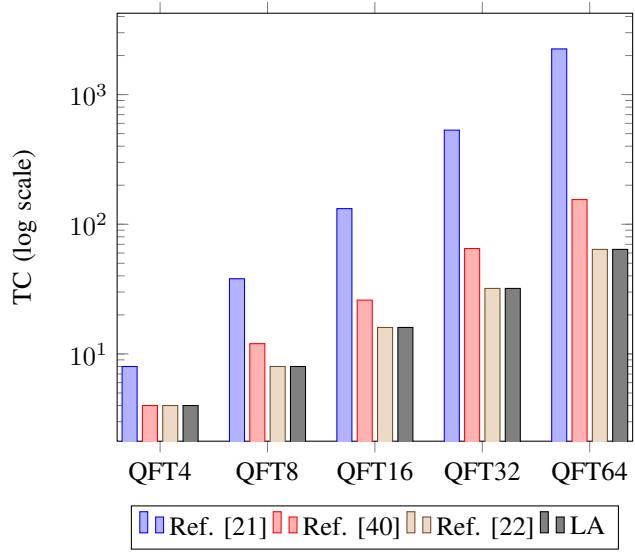

Refs. \cite{houshmand2020evolutionary,chen2023routing} consider the transmission cost within distributed quantum circuits across two partitions. The comparison of transmission cost under two partitions is shown in Fig. \ref{fig:bar_chart_log}. Given the significant differences in transmission cost data across the references, Fig. \ref{fig:bar_chart_log} utilizes a logarithmic axis for the bar chart to minimize the relative differences between data points, making the overall distribution clearer. It is evident that \cite{daei2021improving} and the LA method  demonstrate noticeable optimization effects on the transmission cost for QFT circuits across two partitions, with both achieving lower transmission cost compared to other comparative references.

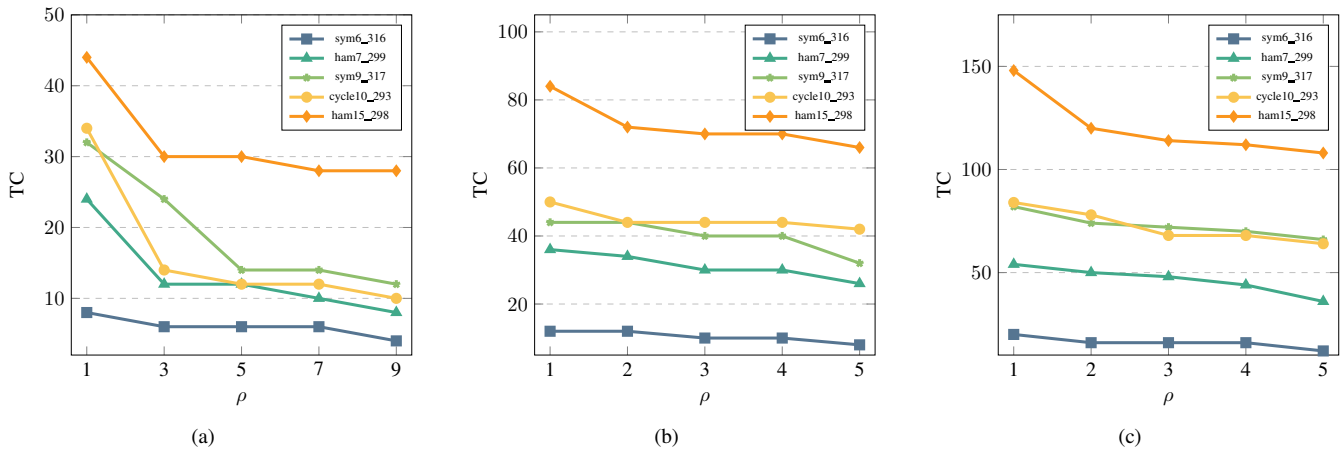
\begin{figure*}[]
	\centering
	\setcounter{subfigure}{0}
	\subfigure[]{
		\begin{adjustbox}{width=0.3\textwidth}  
		\begin{tikzpicture}
			\begin{axis}
				[width=7.5cm, height=7.5cm,   
				line width=0.09pt , 
				xlabel={$\rho$},   
				ylabel={TC},  
				xmin=0, 
				xmax=5,    
				ymin=2, 
				ymax=50, 
				xtick={0,1,2,3,4,5},     
				xticklabels={0,1,2,3,4,5},    
				legend pos=north east,     
				ymajorgrids=true,       
				grid style=dashed,       
				legend style={font=\fontsize{6}{12}\selectfont}, 
				enlarge x limits=0.05,  
                    ylabel style={yshift=-8pt},
				]
				\addplot[color=myblue, line width=1.5pt, mark=square*, mark options={fill=myblue}] 
				coordinates {
					(0,8)(1,6)(2,6)(3,6)(4,4)(5,4)
				};
				\addplot[color=mygreen, line width=1.5pt,mark=triangle*,mark options={fill=mygreen}]
				coordinates {
					(0,24)(1,12)(2,12)(3,10)(4,8)(5,8)
				};
				\addplot[color=myred, line width=1.5pt,mark=star,mark options={fill=myred}]
				coordinates {
					(0,32)(1,24)(2,14)(3,14)(4,12)(5,10)
				};
				\addplot[color=mypurple, line width=1.5pt,mark=otimes*,mark options={fill=mypurple}]
				coordinates {
					(0,34)(1,14)(2,12)(3,12)(4,10)(5,6)
				};
				\addplot [color=myorange,line width=1.5pt, mark=diamond*,mark options={fill=myorange}]
				coordinates {
					(0,44)(1,30)(2,30)(3,28)(4,28)(5,24)
				};
				\legend{sym6\_316, ham7\_299, sym9\_317, cycle10\_293, ham15\_298} 
			\end{axis}
		\end{tikzpicture}
	\end{adjustbox}
 \label{fig6a}
	}
	\quad
	\subfigure[]{  
		\begin{adjustbox}{width=0.3\textwidth}  
		\begin{tikzpicture}
			\begin{axis}
				[width=7.5cm, height=7.5cm,   
				line width=0.09pt , 
				xlabel={$\rho$},   
				ylabel={TC},  
				xmin=0, 
				xmax=5,    
				ymin=5, 
				ymax=105, 
				xtick={0,1,2,3,4,5},     
				xticklabels={0,1,2,3,4,5},    
				legend pos=north east,     
				ymajorgrids=true,       
				grid style=dashed,       
				legend style={font=\fontsize{6}{12}\selectfont}, 
				legend style={row sep=0.001em},
				enlarge x limits=0.05,  
                    ylabel style={yshift=-8pt},
				]
				\addplot[color=myblue, line width=1.5pt, mark=square*, mark options={fill=myblue}] 
				coordinates {
					(0,12)(1,10)(2,8)
				};
				\addplot[color=mygreen, line width=1.5pt,mark=triangle*,mark options={fill=mygreen}]
				coordinates {
					(0,36)(1,34)(2,30)(3,22)(4,18)(5,14)
				};
				\addplot[color=myred, line width=1.5pt,mark=star,mark options={fill=myred}]
				coordinates {
					(0,44)(1,44)(2,40)(3,36)(4,26)(5,22)
				};
				\addplot[color=mypurple, line width=1.5pt,mark=otimes*,mark options={fill=mypurple}]
				coordinates {
					(0,50)(1,44)(2,44)(3,38)(4,34)(5,26)
				};
				\addplot [color=myorange,line width=1.5pt, mark=diamond*,mark options={fill=myorange}]
				coordinates {
					(0,84)(1,72)(2,70)(3,62)(4,52)(5,48)
				};
				\legend{sym6\_316, ham7\_299, sym9\_317, cycle10\_293, ham15\_298} 
			\end{axis}
		\end{tikzpicture}
	\end{adjustbox}
 \label{fig6b}
	}
	\quad
	\subfigure[]{  
		\begin{adjustbox}{width=0.3\textwidth}  
		\begin{tikzpicture}
			\begin{axis}
				[width=7.5cm, height=7.5cm,   
				line width=0.09pt , 
				xlabel={$\rho$},   
				ylabel={TC},  
				xmin=0, 
				xmax=5,    
				ymin=10, 
				ymax=175, 
				xtick={0,1,2,3,4,5},     
				xticklabels={0,1,2,3,4,5},    
				legend pos=north east,     
				ymajorgrids=true,       
				grid style=dashed,       
				legend style={font=\fontsize{6}{12}\selectfont}, 
				legend style={row sep=0.001em},
				enlarge x limits=0.05,  
                    ylabel style={yshift=-8pt},
				]
				\addplot[color=myblue, line width=1.5pt, mark=square*, mark options={fill=myblue}] 
				coordinates {
					(0,20)(1,16)
				};
				\addplot[color=mygreen, line width=1.5pt,mark=triangle*,mark options={fill=mygreen}]
				coordinates {
					(0,54)(1,48)(2,36)(3,32)
				};
				\addplot[color=myred, line width=1.5pt,mark=star,mark options={fill=myred}]
				coordinates {
					(0,82)(1,72)(2,66)(3,58)(4,52)
				};
				\addplot[color=mypurple, line width=1.5pt,mark=otimes*,mark options={fill=mypurple}]
				coordinates {
					(0,84)(1,68)(2,64)(3,62)(4,48)(5,46)
				};
				\addplot [color=myorange,line width=1.5pt, mark=diamond*,mark options={fill=myorange}]
				coordinates {
					(0,148)(1,114)(2,108)(3,94)(4,82)(5,80)
				};
				\legend{sym6\_316, ham7\_299, sym9\_317, cycle10\_293, ham15\_298} 
			\end{axis}
		\end{tikzpicture}
	\end{adjustbox}
 \label{fig6c}
	}
 \caption{Impact of load balancing tolerance \(\rho\) on transmission cost TC under different partitions. (a) (b) (c) shows the TC corresponding to different \(\rho\) under two partitions, three partitions, and four partitions, respectively.}
 \label{fig6}

\end{figure*}

In the above experiments, while optimizing transmission cost, the load balancing tolerance \(\rho\) is set to 0, ensuring the qubit count difference in each partition did not exceed 1. To compare the impact of different \(\rho\) on transmission cost during quantum circuit partitioning, five benchmark circuits are selected. The transmission cost \(TC\) for these circuits is calculated under \(\rho\) values of 1, 3, 5, 7, and 9, depicted in Fig. \ref{fig6}. In a two-partition scenario, with a higher qubit count in subcircuits, the range of \(\rho\) is set from 1 to 9. However, in three or four partition setups, where subcircuits contained fewer qubits, \(\rho\) ranged from 1 to 5.

As load balancing tolerance \(\rho\) increases, transmission cost \(TC\) decreases due to more relaxed load balancing, offering a wider range of partitioning options. This flexibility facilitates finding partitioning methods that minimize global gates and transmission cost. Thus, the selection of \(\rho\) should consider the actual size of the quantum circuit and the distributed execution environment.

\section{Conclusion}\label{sec:6}
In addressing the challenge of quantum state transmission within distributed quantum circuits, this paper introduces a novel approach for circuit partitioning that leverages the QUBO model, coupled with the lookahead method, to efficaciously minimize transmission cost. Furthermore, the integration of the lookahead method enables the selection of optimal transmission qubits within a dynamic lookahead window, effectively reducing the transmission cost of distributed quantum circuits.  Comparative experimental results demonstrate noticeable optimization effects on transmission cost across different numbers of partitions.


%

\section*{Acknowledgments}
The work was supported by the National Natural Science Foundation of China (No. 62072259), in part by the Natural Science Foundation of Jiangsu Province (No. BK20221411), in part by the PhD Start-up Fund of Nantong University (No. 23B03).




%

\bibliographystyle{IEEEtran}
\bibliography{IEEEabrv,IEEEexample}

%

\vfill

\end{document}